\documentclass[aps, prd, twocolumn,showpacs, showkeys, preprintnumbers, nofootinbib, superscriptaddress]{revtex4-2}
\usepackage[sort&compress]{natbib}
\usepackage{multirow}
\usepackage{graphicx}
\usepackage{amssymb,amsbsy,amsmath,amsfonts,longtable}
\usepackage{slashed}
\usepackage{soul}
\usepackage[latin1]{inputenc}

\begin{document}
\title{Bolstering up the existence of $P_s(2080)$}
\author{Breno Agat\~ao}
\email{bgarcia@if.usp.br}
\affiliation{Universidade de Sao Paulo, Instituto de Fisica, C.P. 05389-970, Sao 
Paulo, Brazil.}

\author{A. Vertel Nieto}
\email{anieto@if.usp.br}
\affiliation{Universidade de Sao Paulo, Instituto de Fisica, C.P. 05389-970, Sao 
Paulo, Brazil.}

\author{K. P. Khemchandani}
\email{kanchan.khemchandani@unifesp.br}
\affiliation{Universidade Federal de Sao Paulo, C.P. 01302-907, Sao Paulo, Brazil.}

\author{A. Mart\'inez Torres}
\email{amartine@if.usp.br}
\affiliation{Universidade de Sao Paulo, Instituto de Fisica, C.P. 05389-970, Sao 
Paulo, Brazil.}

\author{Seung-il Nam}
\email{sinam@pknu.ac.kr}
\affiliation{Department of Physics, Pukyong National University (PKNU), Busan 48513, Korea.}
\affiliation{Asia Pacific Center for Theoretical Physics (APCTP), Pohang 37673, Korea.}

\begin{abstract}
We present a detailed study of the partial decay widths of a spin-parity resonance $J^P=3/2^-$ $N^*$ with a mass of $\simeq$ 2070 MeV obtained from the coupled channel s wave vector-baryon $\rho N$, $\omega N$, $\phi N$, $K^*\Lambda$ and $K^*\Sigma$ dynamics.  This state, which couples strongly to the $K^*\Sigma$ channel, corresponds to a nucleon with a hidden strange quark content, in analogy to the $P_c$ states discovered by the LHCb collaboration, and we denote it as $P_s(2080)$. A state with such a nature can decay to vector-baryon, pseudoscalar-baryon, and pseudoscalar-baryon resonance channels, involving triangular loops in the latter two cases. As we will show, the partial decay widths to pseudoscalar-baryon resonance channels, like $\pi N^*(1535)$, $\pi N^*(1650)$, $K\Lambda(1405)$, are comparable to those related to ground state baryons in the final state, like $\pi N$, $\eta N$, $K\Lambda$. In this way, reactions involving such lighter baryon resonances in the final state can be used as an alternative source of information on the properties of a $N^*$ with hidden strangeness.
\end{abstract}



\maketitle
\date{\today}

\section{Introduction}
The discovery of the $P_c$ pentaquarks by the LHCb collaboration~\cite{LHCb:2015yax,LHCb:2016ztz,LHCb:2019kea} has undoubtedly proven the existence of exotic baryons
whose properties cannot be understood in terms of three quarks. Their nature and quantum numbers, however, are still unclear, and different spin-parity assignments and inner structures, like pentaquarks or meson-baryon molecular type of hadrons, have been proposed for describing the $P_c$ states~\cite{Maiani:2015vwa,Roca:2015dva,Guo:2015umn,Liu:2015fea,Weng:2019ynv,Liu:2019tjn,Burns:2019iih,He:2019ify,Xiao:2019mvs,Guo:2019kdc,Fernandez-Ramirez:2019koa}. 

\begin{figure}
\includegraphics[width=0.45\textwidth]{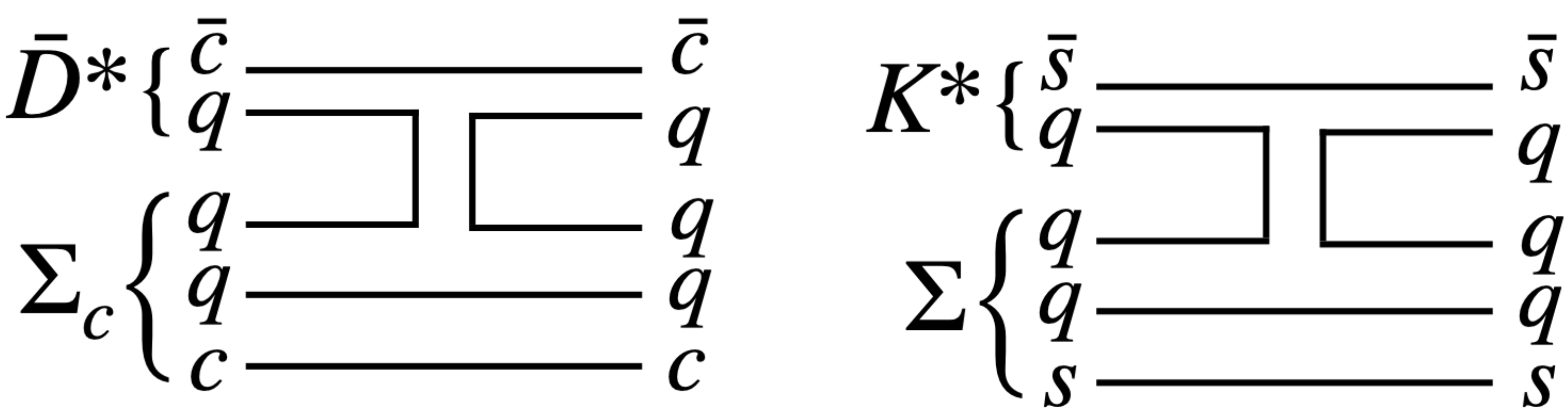}
\caption{(Left) Vector exchange in the $t$-channel for the process $\bar D^* \Sigma_c\to \bar D^*\Sigma_c$ (Right) Same mechanism but for $K^*\Sigma\to K^*\Sigma$. In both cases, the heavy quarks behave as expectators.}\label{quarkdiagram}
\end{figure}

The $P_c$ states, being observed in the $J/\psi$-$p$ invariant mass distribution of the process $\Lambda^0_b\to J/\psi p K^-$, correspond to nucleon resonances with hidden charm and one could wonder if there may exist in Nature their hidden strange partners.  If the $P_c$ states would be generated from the meson-baryon dynamics, $P^+_c(4450)$ seems to be described as a spin-parity $J^P=3/2^-$, isospin $I=1/2$ baryon obtained mainly from the interaction of $\bar D^*$ and $\Sigma_c$ in the $s$-wave, and whose nominal mass is $\simeq 8$ MeV below the threshold of the latter channel. One of the relevant contributions in the description of the $\bar D^*$ and $\Sigma_c$ interaction consists of exchanging a vector meson, like $\rho$, $\omega$, in the $t$-channel~\cite{Wu:2010jy}. In such a case, the quark $\bar c$ in $\bar D^*$ and the quark $c$ in $\Sigma_c$ act as spectators, as shown in Fig.~\ref{quarkdiagram}. If the quarks $\bar c$ and $c$ are now replaced by the quarks $\bar s$ and $s$, respectively, the $\bar D^* \Sigma_c$ system would become $K^*\Sigma$, interacting via vector meson exchange in the $t$-channel, with the quarks $\bar s$ and $s$ continuing being spectators as well. Since in both cases, the heavy quarks in the respective systems behave as spectators, assuming the relevant dynamics needed to form states in such systems to be the $t$-channel exchange of vector mesons, the formation of an isospin $1/2$ state, with $J^P=3/2^-$ and a mass of $\simeq 2077$ MeV, i.e., $\simeq$ 8 MeV below the $K^*\Sigma$ threshold, in analogy with $\bar D^*\Sigma_c$, seems almost compelling. 

After the discovery of the $P_c$ states, several authors have investigated the existence of the hidden strange partners of the former. For instance, in Ref.~\cite{He:2017aps},  the $3/2^-$ nucleon resonances $N^*(1875)$ and $N^*(2120)$ were interpreted as hadronic molecular states, generated from the coupled channel interactions $\Sigma^* K$ and $\Sigma K^*$ considering a boson exchange potential model to solve the Bethe-Salpeter equation. In Ref.~\cite{Lin:2018kcc}, by assuming that $N^*(1875)$ and $N^*(2120)$ are indeed $s$-wave  $K\Sigma^*$ and $K^*\Sigma$ states, and by fixing the mass of these states to be, respectively, 1875 and 2080 MeV, the coupling constants of the former resonances to the latter states were determined by considering the Weinberg compositeness condition~\cite{Weinberg:1965zz}. Using the obtained coupling constants and effective Lagrangians to describe the vertices, the partial decay widths of the mentioned $N^*$ resonances to final states formed by vector-baryon, pseudoscalar-baryon and $K\Lambda(1405)$, $K\Lambda(1520)$ were determined by considering a pion exchange in a triangular loop.

The production of hidden strangeness nucleon resonances with a mass of $\sim$ 2000-2100 MeV has been theoretically studied~\cite{Ramos:2013wua,Kiswandhi:2010ub,Xie:2010yk,Xie:2013db,Kiswandhi:2016cav,Gao:2017hya} in the past and related to some of the bump-like structures observed in the experimental data in the same energy region~\cite{LEPS:2005hax,Dey:2014tfa}, in different processes involving final states such as $\phi N$, $K\Sigma$. More recently, within the context of the existence of hidden strangeness partners of the $P_c$ states, the presence of $P_s(2080)$ has been investigated in data on processes such as $\gamma p\to \phi p$, $K^+p\to K^+\phi p$, $\pi^- p\to \phi n$~\cite{Nam:2021ayk,Wu:2023ywu,Wang:2024qnk,Ben:2023uev}. In some of these studies, the existence of such states, the mass, width, and quantum numbers are assumed using the analogy with the $\bar D^*\Sigma_c$ interaction, and the coupling constants needed to evaluate the corresponding cross sections are determined either via the Weinberg compositeness condition, considering the $P_s$ states to be bound states of two hadrons whose threshold is close to the assumed mass, or from fits to the data~\cite{Wu:2023ywu,Wang:2024qnk,Ben:2023uev}. 

Before continuing with further discussions, we should mention that no $N^*$ resonance with a mass of 2080 MeV and quantum numbers $J^P=3/2^-$ is listed in the most recent version of the Review of Particle Physics (PDG)~\cite{ParticleDataGroup:2024cfk}: before the 2012 version of the PDG, any evidence for $N^*$ resonances with $J^P=3/2^-$ and mass above 1800 MeV were collected under the label of $N^*(2080)$. In the latest volume, two $J^P=3/2^-$ states, a three-star $N^*(1875)$ and a two-star $N^*(2120)$, are cataloged. However, a closer look at the papers listed in the PDG in these entries shows a large uncertainty ($\sim$ 100 MeV) in these states' mass and width values. Therefore, it needs to be clarified if a bunch of different states might be listed under the same entry.

Despite the absence of $N^*(2080)$ in the latest Review of Particle Physics, theoretical evidence for its existence and its nature as a $3/2^-$ $K^*\Sigma$ quasibound state was reported in Ref.~\cite{Khemchandani:2011et}, long before the discovery of the $P_c$ states by the LHCb collaboration. In Refs.~\cite{Khemchandani:2011et, Khemchandani:2013nma}, the coupled channel $K^*\Sigma$, $K^*\Lambda$, $\phi N$, $\omega N$ and $\rho N$ vector-baryon dynamics was studied by using effective Lagrangians based on the hidden local symmetry~\cite{Bando:1987br}, considering $t$-, $s$-, $u$-channel exchange contributions as well as a contact interaction whose origin lies in the nature of the Lagrangian considered. The amplitudes were projected on the $s$-wave and further on the spin 1/2 and 3/2 bases. As a consequence of the aforementioned dynamics, the generation of several $J^P=1/2^-$ and $3/2^-$ $N^*$ and $\Delta$ resonances were found, and, in particular, for the case of $J^P=3/2^-$ and isospin 1/2, a pole in the second Riemann sheet with a mass of $\simeq 2071$ MeV and a width\footnote{There is a typo in the original work, in which the full width obtained for the state is referred to as the half-width of the state.} of $\simeq 60-70$ MeV was obtained, with the state having a large coupling to the $K^*\Sigma$ channel. 

Denoting the former state as $N^*(2080)$, given its large coupling to $K^*\Sigma$ and the proximity of its mass to the threshold of this channel, such a state can be considered as a nucleon resonance with hidden strangeness. In analogy to the notation for the $P_c$ states, we could use the nomenclature $P_s(2080)$ to represent the state, where the letter $P$ refers to the five quark (pentaquark) content (four quarks and an anti-quark) and the subscript $s$ to the presence of a $s\bar s$ pair in the inner structure of the state. 

It is worth mentioning that the generation of nucleon resonance with hidden strangeness content, from vector-baryon dynamics, was also investigated in Ref.~\cite{Oset:2010tof}. In this former work, considering $t$-channel exchange contributions, spin-degenerate amplitudes were obtained, which led to the finding of two $N^*$ resonances, both with mass of 1977 MeV, and width of 106 MeV, but different spin-parity quantum numbers (one having $J^P=1/2^-$ and another having $J^P=3/2^-$). 

The study of Ref.~\cite{Oset:2010tof} was revisited in Ref.~\cite{Ramos:2013wua}, where the cross sections for
$\gamma p\to K^0\Sigma^+$, $\gamma n\to K^0\Sigma^0$ were determined and the role of the production of $N^*$ resonances with hidden strangeness near the $K^*\Lambda$ and $K^*\Sigma$ thresholds was studied. By readjusting the model parameters used in Ref.~\cite{Oset:2010tof} to regularize the vector-baryon loops entering the Bethe-Salpeter equation, the pole at $M-i\Gamma/2=1977-i53$ MeV was shifted to $\simeq 2035-i 63$ MeV, providing an interpretation to the bump observed in the cross-section of $\gamma p\to K^0\Sigma^+$ at energies around 2000 MeV (which is close to the $K^*\Lambda$ threshold). In this way,  according to the authors of Ref.~\cite{Ramos:2013wua}, there should be two $N^*$ resonances with hidden strangeness at $\simeq 2035-i63$ MeV, one with $J^P=1/2^-$ and other with $J^P=3/2^-$. 

The findings of Refs.~\cite{Khemchandani:2011et, Khemchandani:2013nma} are different to those of Ref.~\cite{Oset:2010tof}. In Refs.~\cite{Khemchandani:2011et, Khemchandani:2013nma} $N^*$ resonances with different masses for $J^P=1/2^-$ and  $J^P=3/2^-$ were obtained in the energy region of $\sim$ 1900-2100~MeV. In $J^P=1/2^-$ two overlapping poles were found at $1801-i 96$ and $1912-i 54$ MeV, which produce one peak on the real axis and were related to $N^*(1895)$. Such a nature of $N^*(1895)$ was found to be useful in describing the cross sections for $\gamma p\to K^+\Lambda(1405)$~\cite{Kim:2021wov}. The $J^P=3/2^-$ state, obtained at $\simeq 2071-i 35$ MeV, was related to the $J^P=3/2^-$ $N^*(2080)$ appearing in the previous version of the PDG.

To summarize this discussion, we can say that there seems to gather evidences for the existence of a $3/2^-$ state with mass around 2080 MeV in recent times. Some works assume such a possibility and search for the signals of a hidden strange partner of $P_c(4457)$ in the experimental data, and in some works a simplified model is used to determine meson-baryon scattering amplitudes. The experimental data too are still scarce to draw clear conclusions. Here, we benefit from the work of Ref.~\cite{Khemchandani:2011et} which, using a more complete framework, predicted the existence of $N^*(2080)$, and study its decay to channels like $\pi N$, $\eta N$, $K\Lambda$. We also explore decay channels involving baryon resonances, such as $\pi N^*(1535)$, $\pi N^*(1650)$, $\eta N^*(1535)$, $K\Lambda(1405)$, which could serve as alternative processes to search for a $P_s$-state, i.e., a non-strange partner of the $\bar D^*\Sigma$ quasibound state.

\section{Calculation of the partial decay widths}
\begin{figure}[h!]
\centering
\includegraphics[width=0.45\textwidth]{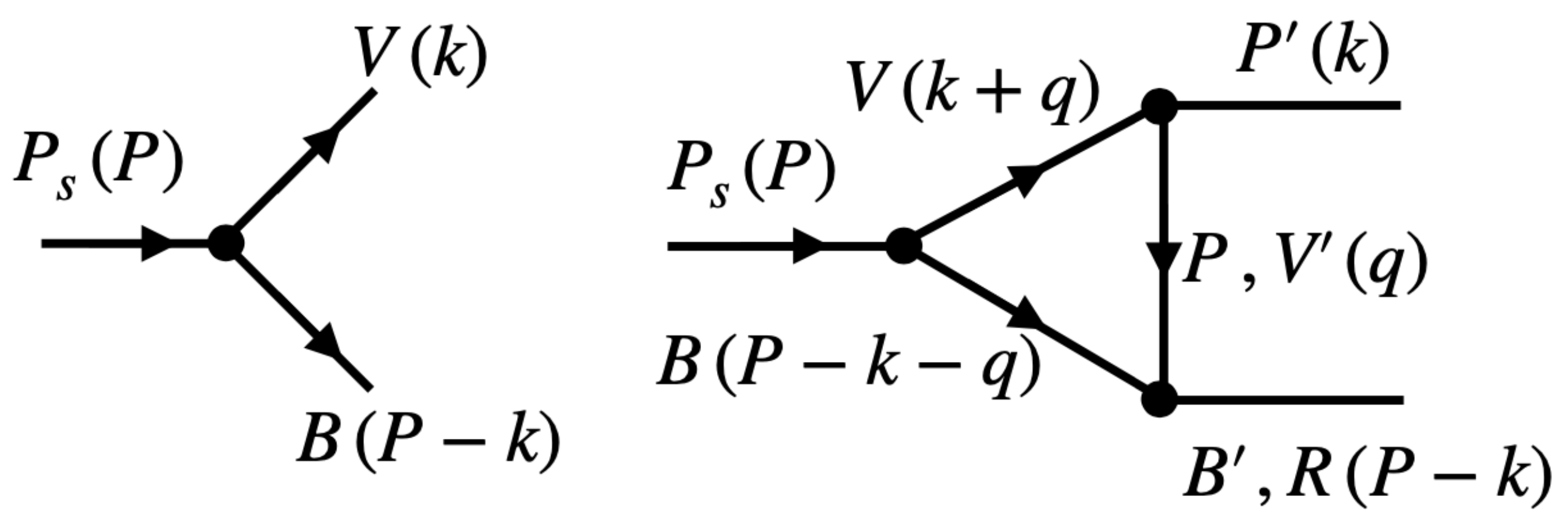}
\caption{Decay mechanisms for $P_s(2080)$ [$J^P=3/2^-$] to VB (Left) and to PB/PR (Right) channels, where $R$ represents either $N^*(1535)$, $N^*(1650)$ or $\Lambda(1405)$, which are $J^P=1/2^-$ states. The four-momenta assignment for each particle is shown between brackets.}\label{Psdecay}
\end{figure}

We start the discussions by showing in Fig.~\ref{Psdecay} different decay mechanisms for the $P_s(2080)$ found in Ref.~\cite{Khemchandani:2011et}. Since the former state is obtained from the $s$-wave vector-octet baryon coupled channel (VB) dynamics with $J^P=3/2^-$, we can have a direct decay mode of $P_s(2080)$ to the VB channels considered for its generation: $K^*\Sigma$, $K^*\Lambda$, $\phi N$, $\omega N$, and $\rho N$. In this case, the amplitude describing such a process can be written as
\begin{align}
-it_{P_s\to V_iB_i}=ig_{P_s\to V_iB_i}\bar u_{B_i}(P-k)\epsilon^\mu_{V_i}(k)u_{P_s\mu}(P),\label{tPs}
\end{align}
where $g_{P_s\to V_iB_i}$ represents the coupling constant of $P_s(2080)$ to a VB channel $i$ constituted by a vector $V_i$ and a baryon $B_i$, $\epsilon^\mu_{V_i}$ is the polarization vector associated with the vector meson $V_i$, $u_{P_s\mu}$ is a Rarita-Schwinger spinor~\cite{Rarita:1941mf} related to $P_s$, and $P^\mu$, $k^\mu$ represent the four-momenta of $P_s$ and of the meson in the final state, respectively. To simplify the notation, the dependence of the spinors on the spin projection of the corresponding particle has been omitted in Eq.~(\ref{tPs}). The Dirac and Rarita-Schwinger spinors related, respectively, to particles of four-momenta $Q$, masses $m$ and $M$ and spin projections $\alpha$ and $\beta$, are normalized such that~\cite{Wu:2023ywu,Xie:2010yk}
\begin{align}
&\sum\limits_{\alpha=-1/2}^{1/2} u(Q,\alpha)\bar u(Q,\alpha)=\frac{(\slashed{Q}+m\mathbb{I})}{2m},\nonumber\\
&\sum\limits_{\beta=-3/2}^{3/2} u_\mu(Q,\beta)\bar u_\nu (Q,\beta)=\frac{\slashed{Q}+M\mathbb{I}}{2M}P_{\mu\nu},
\end{align}
where
\begin{align}
&P_{\mu\nu}=-g_{\mu\nu}\mathbb{I}+\frac{1}{3}\gamma_\mu\gamma_\nu+\frac{2}{3}\frac{Q_\mu Q_\nu}{M^2}\mathbb{I}+\frac{\gamma_\mu Q_\nu-\gamma_\nu Q_\mu}{3M},\nonumber
\end{align}
with $\mathbb{I}$ being the identity matrix.

When considering the process $V_iB_i\to P_s\to V_jB_j$ in the $s$-wave, the amplitude in Eq.~(\ref{tPs}) gives rise to the following isospin $1/2$, $s$-wave, and spin $3/2$ projected amplitude $T^{S=3/2}_{ij}$, i.e., $I(J^P)=1/2\,(3/2^-)$, in the non-relativistic limit, 
\begin{align}
T^{S=3/2}_{ij}(\sqrt{s})=\frac{g_{P_s\to V_iB_i}g_{P_s\to V_jB_j}}{\sqrt{s}-m_{P_s}+i\Gamma_{P_s}/2},\label{TS}
\end{align}
with $m_{P_s}$ ($\Gamma_{P_s}$) being the mass (width) of $P_s(2080)$ and $\sqrt{s}$ representing the center-of-mass energy of the system. Equation (\ref{TS}) shows that the coupling constants $g_{P_s\to V_iB_i}$ needed in Eq.~(\ref{tPs}) can be directly obtained from the residue of the $t$-matrix describing the $V_iB_i\to V_iB_i$ interaction in which $P_s$ is dynamically generated. 
These coupling constants were determined in Ref.~\cite{Khemchandani:2011et} from the analytical continuation of the $t$-matrix in the second Riemann sheet. Alternatively, it is possible to calculate the mentioned coupling constants by using the $t$-matrix determined in Ref.~\cite{Khemchandani:2011et} on the real-energy plane. In this case, from Eq.~(\ref{TS}), we can calculate the coupling constants $g_{P_s\to V_iB_i}$ as
\begin{align}
g_{P_s\to V_iB_i}=\sqrt{i\frac{\Gamma_{P_s}}{2}T^{S=3/2}_{ii}(m_{P_s})}.
\end{align}
For a channel $j\neq i$, the couplings $g_{P_s\to V_jB_j}$ are obtained from the ratio between $T^{S=3/2}_{ij}$ and $T^{S=3/2}_{ii}$ at $\sqrt{s}=m_{P_s}$, i.e., 
\begin{align}
g_{P_s\to V_jB_j}=g_{P_s\to V_iB_i}\frac{T^{S=3/2}_{ij}(m_{P_s})}{T^{S=3/2}_{ii}(m_{P_s})}.
\end{align}
In this way, all the relative phases between the couplings for $i\neq j$ are all related to the same channel $i$. This latter procedure of calculating the coupling constants is more convenient when considering the finite width of the $\rho$- and $K^*$-mesons through a convolution of the loop functions while solving the Bethe-Salpeter equation for the coupled channel system. Here, we follow the latter approach and provide the obtained coupling constants in Table~\ref{couplings}. As can be seen in the table, the coupling of $P_s(2080)$ to the $K^*\Sigma$ channel, whose nominal threshold (2085 MeV) is the closest to the mass of $P_s$, is the largest, as implicitly assumed in Refs.~\cite{Wu:2023ywu,Wang:2024qnk} when considering the Weinberg compositeness condition to determine the coupling constant of $P_s(2080)$ to $K^*\Sigma$ by considering $P_s$ to be a $K^*\Sigma$ bound state. However, this does not necessarily mean that the other coupled channels listed in Table~\ref{couplings} will have no relevant contributions to the partial decay widths of $P_s$, especially when considering the triangular loop mechanisms shown in Fig.~\ref{Psdecay}, where the interference effects between different coupled channels can play a relevant role obtaining the partial decay widths.

\begin{table}[h!]
\centering
\caption{Coupling constants (dimensionless) of $P_s(2080)$ to the vector-baryon channels, in the isospin $1/2$ basis, considered for its generation.}\label{couplings}
\begin{tabular}{c|ccc}\\
Channel&$\rho N$&$\omega N$&$\phi N$\\
Coupling&$-0.231-i0.284$&$-0.175+i0.038$&$0.285+i0.01$\\\hline
Channel&$K^*\Lambda$&$K^*\Sigma$&\\
Coupling&$0.112+i0.553$&$2.313-i0.856$
\end{tabular}
\end{table}

With the coupling constants listed in Table~\ref{couplings}, and considering the rest frame of the decaying particle, the amplitudes in Eq.~(\ref{tPs}) can be evaluated and the partial decay width of $P_s(2080)$ to a $V_iB_i$ channel can be determined from
\begin{align}
&\Gamma_{P_s\to V_i B_i}(m_{P_s},m_{V_i},m_{B_i})=\frac{m_{B_i}}{m_{P_s}}\frac{|\pmb{p}_{i}|}{(2\pi)}\frac{1}{2s_{P_s}+1}\nonumber\\
&\quad\times\sum\limits_{\text{pol.}}|t_{P_s\to V_i B_i}|^2\Theta(m_{P_s}-m_{V_i}-m_{B_i}),\label{width}
\end{align}
where $|\pmb{p}_{i}|$ is the modulus of the center-of-mass linear momentum of the particles in the final state, $s_{P_s}$ is the spin of $P_s(2080)$, $\Theta(\cdots)$ is the Heavisde $\Theta$-function, and the symbol $\sum\limits_\text{pol.}$ represents summing over the polarizations of the particles in the initial and final states. The finite width of $P_s$  can be incorporated by considering a convolution of the expression in Eq.~(\ref{width}) with the corresponding spectral function for $P_s$:
\begin{align}
\Gamma_{P_s\to V_i B_i}&=\frac{1}{N_{P_s}}\int\limits_{m_{P_s}-2\Gamma_{P_s}}^{m_{P_s}+2\Gamma_{P_s}} d\tilde{m}_{P_s}\rho_{P_s}(\tilde{m}_{P_s})\nonumber\\
&\quad\times\Gamma_{P_s\to V_i B_i}(\tilde{m}_{P_s},m_{V_i},m_{B_i}),
\end{align}
where 
\begin{align}
\rho_{P_s}(\tilde{m}_{P_s})&=-\frac{1}{\pi}\text{Im}\Bigg(\frac{1}{\tilde{m}_{P_s}-m_{P_s}+i\Gamma_{P_s}/2}\Bigg),\label{Spec}
\end{align}
and $N_{P_s}$ is the normalization of the spectral function of Eq.~(\ref{Spec}) when considering $\tilde{m}_{P_s}\in[m_{P_s}-2\Gamma_{P_s},m_{P_s}+2\Gamma_{P_s}]$, 
\begin{align}
N_{P_s}=\int\limits_{m_{P_s}-2\Gamma_{P_s}}^{m_{P_s}+2\Gamma_{P_s}}d\tilde{m}_{P_s}\rho_{P_s}(\tilde{m}_{P_s}).
\end{align}
Note that the effect of the finite width of the vector mesons in the final state is already present in the coupling constants listed in Table~\ref{couplings}.

In the case of $P_s$ decaying to pseudoscalar-baryon ($P^\prime_i B_i$) or pseudoscalar-baryon resonance ($P^\prime_i R_i$, with $R_i$ having $J^P=1/2^-$) channels, the decay mechanism proceeds via triangular loops, as shown in Fig.~\ref{Psdecay}. Now we can have contributions from the exchange of pseudoscalars ($P$)  or vector mesons ($V^\prime$) between the vectors ($V$) and baryons ($B$) produced in the primary vertex. For instance, we can have channels in the final state like $\pi N$, $\eta N$, $K\Lambda$, $K\Sigma$, $\eta^\prime N$, $K\Lambda(1405)$, $\pi N^*(1535)$, $\eta N^*(1535)$ or $\pi N^*(1650)$, and intermediate states in the triangular loop like $K^*\Sigma\pi$, $K^*\Sigma\eta$, $K^*\Sigma\eta^\prime$, $K\Lambda\pi$, $\rho N \bar K$, $\omega N\bar K$, $\phi N\bar K$, $K^*\Sigma\omega$, etc. Thus, to evaluate the contribution to the partial decay widths of $P_s$ from the diagrams represented in Fig.~\ref{Psdecay}, we need amplitudes describing the vector-pseudoscalar-pseudoscalar (VPP), vector-vector-pseudoscalar (VVP), pseudoscalar-baryon-baryon (PBB), and vector-baryon-baryon (VBB) vertices.  These latter contributions are determined from effective Lagrangians based on the chiral and hidden local symmetries~\cite{Ecker:1994gg,Bernard:1995dp,Bando:1987br}, with
\begin{align}
\mathcal{L}_{VPP}&=-i g\langle V^\mu[P,\partial_\mu P]\rangle,\nonumber\\
\mathcal{L}_{VVP}&=\frac{G}{\sqrt{2}}\epsilon^{\mu\nu\alpha\beta}\langle \partial_\mu V_\nu\partial_\alpha V_\beta P\rangle,\nonumber\\
\mathcal{L}_{PBB}&=-\frac{D+F}{\sqrt{2}f_\pi}\langle \bar B\gamma^\mu\gamma_5\partial_\mu P B\rangle-\frac{D-F}{\sqrt{2}f_\pi}\langle \bar B\gamma^\mu\gamma_5 B\partial_\mu P\rangle,\nonumber\\
\mathcal{L}_{VBB}&=g[\langle \bar B\gamma_\mu[V^\mu,B]\rangle+\langle \bar B\gamma_\mu B\rangle \langle V^\mu\rangle],\label{Leff}
\end{align}
where $g=m_V/(2 f_\pi)$, $m_V\simeq m_\rho$, $G=3g^2/(4\pi^2 f_\pi)$, $D\simeq 0.80$, $F\simeq0.46$, $f_\pi\simeq 93$ MeV, $u=e^{i P/(\sqrt{2} f_\pi)}$, $u_\mu=iu^\dagger \partial_\mu U u^\dagger$, $U=u^2$, $P$, $B$, and $V^\mu$ are matrices whose elements are, respectively, the pseudoscalar, baryon and vector fields from the octet,
\begin{align}
P&=\left(\begin{array}{ccc}\frac{\eta}{\sqrt{3}}+\frac{\eta^\prime}{\sqrt{6}}+\frac{\pi^0}{\sqrt{2}}&\pi^+&K^+\\\pi^-&\frac{\eta}{\sqrt{3}}+\frac{\eta^\prime}{\sqrt{6}}-\frac{\pi^0}{\sqrt{2}}&K^0\\K^-&\bar K^0&-\frac{\eta}{\sqrt{3}}+\sqrt{\frac{2}{3}}\eta^\prime\end{array}\right),\nonumber\\
B&=\left(\begin{array}{ccc}\frac{\Sigma^0}{\sqrt{2}}+\frac{\Lambda}{\sqrt{6}}&\Sigma^+&p\\\Sigma^-&-\frac{\Sigma^0}{\sqrt{2}}+\frac{\Lambda}{\sqrt{6}}&n\\\Xi^-&\Xi^0&-\frac{2}{\sqrt{6}}\Lambda\end{array}\right),\nonumber\\
V_\mu&=\left(\begin{array}{ccc}\frac{\omega+\rho^0}{\sqrt{2}}&\rho^+&K^{*+}\\\rho^-&\frac{\omega-\rho^0}{\sqrt{2}}&K^{*0}\\K^{*-}&\bar K^{*0}&\phi\end{array}\right)_\mu,\label{PVB}
\end{align}
and the symbol $\langle\cdots\rangle$ indicates SU(3) trace. Here ideal $\eta$-$\eta^\prime$ mixing, i.e., a mixing angle of $\beta\simeq -19.43^\circ$ ($\text{sin}\beta=-1/3$) has been assumed when writing the elements of the matrix $P$. A value of $\beta$ in the range $\simeq -15^\circ$ to$-22^\circ$ is compatible with the experimental data~\cite{Gilman:1987ax,Akhoury:1987ed,Bramon:1997mf}, and such uncertainty will be considered in the calculation of the partial decay widths. The expression of $P$ in terms of a general mixing angle $\beta$ can be found in appendix~\ref{coeff}. 

In the case of a $J^P=1/2^-$ baryon resonance in the final state, we consider the amplitudes~\cite{Kim:2021wov}
\begin{align}
-it_{PB\to R}&=i g_{R\to PB}\bar u_R(p)u_B(P-k-q),\nonumber\\
-i t_{V^\prime B\to R}&=-\frac{g_{R\to V^\prime B}}{\sqrt{3}}\epsilon^\mu_{V^\prime}(q)\bar u_R(p)\gamma_\mu\gamma_5 u_B(P-k-q),\label{tR}
\end{align}
with $g_{R\to PB\,(V^\prime B)}$ being the coupling constant of the resonance $R$ to the PB and VB channels considered for its generation and $u_R$ being the Dirac-spinor related to the $J^P=1/2^-$ baryon in the final state. The factor $1/\sqrt{3}$ in Eq.~(\ref{tR}) has its origin in the fact that the $g_{R\to V^\prime B}$ coupling in Refs.~\cite{Khemchandani:2013nma,Khemchandani:2011mf,Khemchandani:2018amu} are determined by parametirizing the meson-baryon $t$-matrices as Breit-Wigner amplitudes while Eq.~(\ref{tR}) provides a spin dependent expression (see Ref.~\cite{Kim:2021wov} for more details). Here we consider the low-lying $\Lambda$ and $N^*$ resonances for which phase space is available for decaying, i.e., $\Lambda(1405)$, $N^*(1535)$ and $N^*(1650)$, and use the coupling constants determined in Refs.~\cite{Khemchandani:2011mf,Khemchandani:2013nma,Khemchandani:2018amu}, where PB and VB channels were treated as coupled channels when solving the Bethe-Salpeter equations and the couplings constants were determined from the residues of the corresponding $T$-matrix in the complex energy plane. 

In the case of the process $P_s\to P^\prime B^\prime$ shown in Fig.~\ref{Psdecay}, using the previous amplitudes and the effective Lagrangians in Eq.~(\ref{Leff}), we get the following contribution for a particular vector-baryon-pseudoscalar (VBP) channel in the triangular loop shown in Fig.~\ref{Psdecay}:
\begin{align}
&-i t^{VBP}_{P_s\to P^\prime B^\prime}=g C_{PB\to B^\prime}C_{V\to P^\prime P}g_{P_s\to VB}\bar u_{B^\prime}(p)\gamma_5\nonumber\\
&\quad\times\Bigg[(2 p^\nu+(m_B+m_{B^\prime})\gamma^\nu)(-I^{(1)}_{\nu\mu}+k_\mu I^{(2)}_\nu)\nonumber\\
&\quad+I^{(3)}_{\mu}-k_\mu I^{(4)}\Bigg]u^\mu_{P_s}(P),\label{tVBP}
\end{align}
where $C_{PB\to B^\prime}$ and $C_{V\to P^\prime P}$ are coefficients obtained from the effective Lagrangians of Eq.~(\ref{Leff}) and 
\begin{align}
I^{(1)}_{\nu\mu}&=\Big(1+\frac{k^2}{m^2_V}\Big)\mathbb{I}^{(1)}_{\nu\mu}-\frac{1}{m^2_V}\mathbb{I}^{(2)}_{\nu\mu},\nonumber\\
I^{(2)}_\nu&=\Big(1-\frac{k^2}{m^2_V}\Big)\mathbb{I}^{(3)}_\nu+\frac{1}{m^2_V}\mathbb{I}^{(4)}_\nu,\nonumber\\
I^{(3)}_\mu&=\Big(1+\frac{k^2}{m^2_V}\Big)\mathbb{I}^{(4)}_\mu-\frac{1}{m^2_V}\mathbb{I}^{(5)}_\mu,\nonumber\\
I^{(4)}&=\Big(1-\frac{k^2}{m^2_V}\Big)\mathbb{I}^{(6)}_\nu+\frac{1}{m^2_V}\mathbb{I}^{(7)},
\end{align}
with
\begin{align}
\mathbb{I}^{(1)}_{\nu\mu}&=\int\frac{d^4q}{(2\pi)^4}\frac{q_\nu q_\mu}{\mathbb{D}};~
\mathbb{I}^{(2)}_{\nu\mu}=\int\frac{d^4q}{(2\pi)^4}\frac{q^2q_\nu q_\mu}{\mathbb{D}};\nonumber\\
\mathbb{I}^{(3)}_{\nu}&=\int\frac{d^4q}{(2\pi)^4}\frac{q_\nu }{\mathbb{D}};~
\mathbb{I}^{(4)}_{\nu}=\int\frac{d^4q}{(2\pi)^4}\frac{q^2 q_\nu}{\mathbb{D}},\nonumber\\
\mathbb{I}^{(5)}_{\mu}&=\int\frac{d^4q}{(2\pi)^4}\frac{q^4q_\mu}{\mathbb{D}};~
\mathbb{I}^{(6)}=g^{\nu\mu}\mathbb{I}^{(1)}_{\nu\mu};\nonumber\\
\mathbb{I}^{(7)}&=g^{\nu\mu}\mathbb{I}^{(2)}_{\nu\mu};\nonumber\\
\mathbb{D}&=[(P-k-q)^2-m^2_B+i\epsilon][(k+q)^2-m^2_V+i\epsilon]\nonumber\\
&\quad\times[q^2-m^2_P+i\epsilon].\label{Itensor}
\end{align}
It should be noted that the expressions in Eqs.~(\ref{tR})-(\ref{Itensor}) depend on the particular channel considered in the final and intermediate states, but to simplify the notation we omit writing such a dependence. 

The integrals in Eq.~(\ref{Itensor}) can be written as combinations of the four-momenta $P_\alpha$ and $k_\beta$ by using Lorentz covariance. For example, after integrating in $d^4q$, $\mathbb{I}^{(1)}_{\mu\nu}$ must be a symmetric tensor of order 2 depending on the four-momenta $P$ and $k$. Thus, we can write
\begin{align}
\mathbb{I}^{(1)}_{\nu\mu}=a^{(1)}_1 g_{\nu\mu}+a^{(1)}_2P_\nu P_\mu+a^{(1)}_3 k_\nu k_\mu+a^{(1)}_4(P_\nu k_\mu+P_\mu k_\nu),
\end{align}
where $a^{(1)}_j$, $j=1,2,\dots,4$ are the coefficients of the  combinations, and which need to be determined. Similar arguments can be used for the other tensor integrals. Details on the calculation of the $a^{(i)}_j$, $i=1,2,\dots,5$, coefficients (which depend on the final and intermediate states) can be found in the appendix~\ref{TI}. The main steps to follow are to use the Passarino-Veltman decomposition of tensor integrals~\cite{Passarino:1978jh}, then determine the $dq^0$ integration analytically by using Cauchy's theorem and the $d^3q$ integration numerically, by using a cut-off or form factors to regularize it. We have varied the cut-off in the range $600-850$ MeV and considered three types of form factors at the vertices (Gaussian, Lorentz, and a Heavise theta-function), and estimated uncertainties in the results 

In terms of the $a^{(i)}_j$ coefficients, the amplitude for the process $P_s\to P^\prime B^\prime$, considering the different $VPB$ intermediate states, thus, exchanging pseudoscalars in the triangular loop of Fig.~\ref{Psdecay}, can be written as
\begin{align}
-it^\text{pseudo}_{P_s\to P^\prime B^\prime}&=-i\sum\limits_{VBP} t^{VBP}_{P_s\to P^\prime B^\prime}\nonumber\\
&=g\bar u_{B^\prime}(p)\gamma_5 t^{(A)}_\mu(\pmb{A},\pmb{\tilde{A}})u^\mu_{P_s}(P),\label{tpseudo}
\end{align}
where
\begin{align}
t^{(A)}_\mu(\pmb{A},\pmb{\tilde{A}})&=\sum\limits_{k=1}^5\big[2 A_k p^\nu+(\tilde A_k+m_{B^\prime}A_k)\gamma^\nu\big]T^{(k)}_{\nu\mu}\nonumber\\
&\quad+A_6P_\mu+A_7 k_\mu. \label{tmu}
\end{align}
In Eq.~(\ref{tmu}), $T^{(k)}_{\nu\mu}$ represents the $k$th-element of $T_{\nu\mu}$, with
\begin{align}
T_{\nu\mu}&=\{g_{\nu\mu},P_\nu P_\mu,k_\nu k_\mu,P_\nu k_\mu,P_\mu k_\nu\},
\end{align}
and $A_i$ and $\tilde{A}_i$ are coefficients given by
\begin{align}
A_i&=\sum\limits_{VBP}C_{PB\to B^\prime}C_{V\to P^\prime P}g_{P_s\to VB}A^{VBP}_i,\nonumber\\
\tilde{A}_i&=\sum\limits_{VBP}C_{PB\to B^\prime}C_{V\to P^\prime P}g_{P_s\to VB}m_BA^{VBP}_i,\label{Ai}
\end{align}
 with $i=1,2,\dots,7$. The $A^{VBP}_i$ coefficients appearing in Eq.~(\ref{Ai}) depend on the four-momenta of the initial and final particles as well as of $a^{(i)}_j$, and their definition can be found in appendix~\ref{TI}.
 
In the case of exchanging a vector ($V^\prime$) between the vector and baryon produced in the primary vertex of the diagram in Fig.~\ref{Psdecay},
 we find the following amplitude describing the process when considering contributions from the different intermediate $VBV^\prime$ channels:
 \begin{align}
 -it^{\text{vector}}_{P_s\to P^\prime B^\prime}&=-i\sum\limits_{VBV^\prime} t^{VBV^\prime}_{P_s\to P^\prime B^\prime}\nonumber\\
 &=-\frac{g G}{\sqrt{2}}\bar u_{B^\prime}(p)t^{(B)}_{\nu^\prime}(\pmb{B})u^{\nu^\prime}_{P_s}(P),\label{tvector1}
 \end{align}
where 
\begin{align}
t^{(B)}_{\nu^\prime}(\pmb{B})&=\epsilon_{\mu^\prime\nu^\prime\alpha^\prime\beta^\prime} k^{\mu^\prime}\bigg[(B_1 g^{\sigma\alpha^\prime}+B_2 P^\sigma P^{\alpha^\prime}\nonumber\\
 &+B_3 P^{\alpha^\prime} k^\sigma)\gamma_\sigma\gamma^{\beta^\prime}+(B_4-m_{B^\prime}B_5)P^{\alpha^\prime}\gamma^{\beta^\prime}\bigg],\nonumber\\
B_i&=\sum\limits_{VBV^\prime}g_{P_s\to VB}C_{V^\prime B\to B^\prime}C_{V\to V^\prime P^\prime}B^{VV^\prime B}_i,\label{Bi}
\end{align}
with $i=1,2,\dots,5$. The $B^{VV^\prime B}_i$ coefficients appearing in Eq.~(\ref{Bi}) are defined in appendix~\ref{TI}. They depend on coefficients, $b^{(i)}_j$, which can be obtained from the expressions for $a^{(i)}_j$ replacing $\omega_P(\pmb{q})$ by $\omega_{V^\prime}(\pmb{q})=\sqrt{\pmb{q}^2+m^2_{V^\prime}}$. The $C_{V^\prime B\to B^\prime}$, and $C_{V\to V^\prime P^\prime}$ in Eq.~(\ref{Bi}) are coefficients obtained from the Lagrangians in Eq.~(\ref{Leff}), and their values can be found in appendix~\ref{coeff}.

Using Eqs.~(\ref{tpseudo}) and (\ref{tvector1}), the sum over the polarizations of the initial and final states for 
\begin{align}
|t_{P_s\to P^\prime B^\prime}|^2=|t^\text{pseudo}_{P_s\to P^\prime B^\prime}+t^\text{vector}_{P_s\to P^\prime B^\prime}|^2
\end{align}
can be calculated, obtaining
\begin{align}
&\sum\limits_{\text{pol.}}|t_{P_s\to P^\prime B^\prime}|^2=\sum\limits_{\text{pol.}}|t^\text{pseudo}_{P_s\to P^\prime B^\prime}|^2+\sum\limits_{\text{pol.}}|t^\text{vector}_{P_s\to P^\prime B^\prime}|^2\nonumber\\
&\quad+2\text{Re}\bigg\{\sum\limits_{\text{pol.}}t^\text{pseudo}_{P_s\to P^\prime B^\prime}\big(t^\text{vector}_{P_s\to P^\prime B^\prime}\big)^\dagger\bigg\},
\end{align}
where
\begin{align}
&\sum\limits_{\text{pol.}}|t^\text{pseudo}_{P_s\to P^\prime B^\prime}|^2=\frac{|g|^2}{4m_{B^\prime}m_{P_s}}\text{Tr}\Bigg[(\slashed{p}-m_{B^\prime})t^{(A)}_\mu(\pmb{A},\pmb{\tilde{A}})\nonumber\\
&\quad\times(\slashed{P}+m_{P_s})P^{\mu\sigma}t^{(A)}_\sigma(\pmb{A}^*,\pmb{\tilde{A}}^*)\Bigg]\nonumber\\
&=\frac{|g|^2}{4m_{B^\prime}m_{P_s}}\sum\limits_{l=0}^5F^{(l)}(P\cdot k)^l,\nonumber\\
&\sum\limits_{\text{pol.}}|t^\text{vector}_{P_s\to P^\prime B^\prime}|^2=\frac{|g|^2|G|^2}{8 m_{B^\prime}m_{P_s}}\text{Tr}\Bigg[(\slashed{p}+m_{B^\prime})t^{(B)}_{\nu^\prime}(\pmb{B})\nonumber\\
&\quad\times(\slashed{P}+m_{P_s})P^{\nu^\prime\sigma^\prime}\tilde{t}^{(B)}_{\sigma^\prime}(\pmb{B}^*)\Bigg]\nonumber\\
&\quad=\frac{|g|^2|G|^2}{8 m_{B^\prime}m_{P_s}}\sum\limits_{l=0}^4H^{(l)}(P\cdot k)^l,\nonumber
\end{align}
\begin{align}
&\sum\limits_{\text{pol.}}t^\text{pseudo}_{P_s\to P^\prime B^\prime}\bigg(t^\text{vector}_{P_s\to P^\prime B^\prime}\bigg)^\dagger\nonumber\\
&\quad=-\frac{|g|^2 G}{4\sqrt{2}m_{B^\prime}m_{P_s}}\text{Tr}\Bigg[(\slashed{p}+m_{B^\prime})\gamma_5 t^{(A)}_\mu(\pmb{A},\pmb{\tilde{A}})\nonumber\\
&\quad\times(\slashed{P}+m_{P_s})P^{\mu\nu^\prime}\tilde{t}^{(B)}_{\nu^\prime}(\pmb{B}^*)\Bigg]\nonumber\\
&\quad=-i\frac{|g|^2 G}{4\sqrt{2} m_{B^\prime}m_{P_s}}\sum\limits_{l=0}^4J^{(l)}(P\cdot k)^l,\label{Tr1}
\end{align}
with
\begin{align}
&\tilde{t}^{(B)}_{\sigma}(\pmb{B}^*)=\epsilon_{\mu^\prime\sigma\alpha^\prime\beta^\prime}k^{\mu^\prime}\bigg[-(B^*_1g^{\sigma^\prime\alpha^\prime}+B^*_2P^{\sigma^\prime}P^{\alpha^\prime}\nonumber\\
&\quad+B^*_3 P^{\alpha^\prime}k^{\sigma^\prime})\gamma_{\sigma^\prime}\gamma^{\beta^\prime}+2B^*_3P^{\alpha^\prime}k^{\beta^\prime}\nonumber\\
&\quad+(B^*_4-m_{B^\prime}B^*_5)P^{\alpha^\prime}\gamma^{\beta^\prime}\bigg].
\end{align}
As can be seen from the preceding equations, the traces present in Eq.~(\ref{Tr1}) can be written as an expansion of powers of $P\cdot k$, with $F^{(l)}$, $H^{(l)}$ and $J^{(l)}$ being the coefficients for such expansions, and their values are given in Appendix~\ref{FHJ}.

Next, we can also have a resonance, like $\Lambda(1405)$, $N^*(1535)$ and $N^*(1650)$, in the final state of a decaying $P_s$ since the former resonances have sizeable couplings to PB~\cite{Kaiser:1995eg,Oset:1997it,Jido:2003cb,Inoue:2001ip,Nieves:2001wt,Bruns:2010sv} and VB channels~\cite{Khemchandani:2011mf,Garzon:2012np,Khemchandani:2013nma,Garzon:2014ida,Khemchandani:2018amu}. We have considered final states formed by a pseudoscalar and one of these resonances, which we denote as $R$. As shown in Fig.~\ref{Psdecay}, the decay mechanism of $P_s\to P^\prime R$ proceeds via triangular loops as well. In this case, the amplitude describing such a process, involving the exchange of pseudoscalar mesons between the hadrons of the primary vertex, can be written as
\begin{align}
t^\text{pseudo}_{P_s\to P^\prime R}=\sum\limits_{VBP}t^{VBP}_{P_s\to P^\prime R}=g\bar{u}_Rt^{(C)}_{\mu}(\pmb{C}) u^\mu_{P_s}(P),\label{tRP}
\end{align}
where
\begin{align}
t^{(C)}_{\mu}(\pmb{C})&=\sum\limits_{k=1}^7 C_k U^{(k)}_\mu,\label{T1mu}
\end{align}
with $U^{(k)}_\mu$ being the $k$th element of
\begin{align}
U_\mu&=\{\gamma^\nu g_{\nu\mu}, \slashed{P} P_\mu, \slashed{k} k_\mu,\slashed{P} k_\mu,\slashed{P} k_\nu,P_\mu,k_\mu\}.
\end{align}
In Eq.~(\ref{T1mu}), $C_k$ are coefficients given by
\begin{align}
C_k=\sum\limits_{VBP} g_{R\to PB}g_{P_s\to VB}C_{V\to P^\prime P}C^{VBP}_i,\label{Ci}
\end{align}
with $i=1,2,\dots,7$, and the definition of $C^{VBP}_i$, which depend on $a^{(i)}_j$, and the four-momenta of the particles in the initial, intermediate and final states, can be found in appendix~\ref{TI}.

In the case of exchanging a vector meson between the particles produced in the primary vertex of the triangular loop for the reaction $P_s\to P^\prime R$, the amplitude can be written, once the contribution from different $VBV^\prime$ channels is included, as
\begin{align}
t^\text{vector}_{P_s\to P^\prime R}=\frac{1}{\sqrt{6}}G\bar u_R(p)t^{(D)}_{\nu^\prime}(\pmb{D})u^{\nu^\prime}_{P_s}(P),\label{tRV}
\end{align}
where
\begin{align}
t^{(D)}_{\nu^\prime}(\pmb{D})&=\epsilon_{\mu^\prime\nu^\prime\alpha^\prime\beta^\prime}k^{\mu^\prime}\gamma_5[(D_1g^{\sigma\alpha^\prime}+D_2P^\sigma P^{\alpha^\prime}\nonumber\\
&\quad+D_3 P^{\alpha^\prime}k^\sigma)\gamma^{\beta^\prime}\gamma_\sigma-D_4 P^{\alpha^\prime}\gamma^{\beta^\prime}],
\end{align}
with
\begin{align}
D_i=\sum\limits_{VBV^\prime}g_{R\to V^\prime B}g_{P_s\to VB}C_{V\to V^\prime P^\prime}D^{VBV^\prime}_i,\label{Di}
\end{align}
where $i=1,2,\dots,4$. We refer the reader to appendix~\ref{TI} for the definition of the coefficients $D^{VBV^\prime}_i$. We should mention at this point that the coupling constants $g_{R\to PB}$ and $g_{R\to V^\prime B}$ can be found, for instance, in Refs.~\cite{Khemchandani:2018amu,Khemchandani:2013nma}. There, $\eta-\eta^\prime$ mixing was not considered, but the coupling constants of $R$ to the channels $\eta B$ ($\eta^\prime B$) can be estimated by multiplying those obtained in Refs.~\cite{Khemchandani:2018amu,Khemchandani:2013nma} by $\text{cos}\beta$ ($\text{sin}\beta$), where the latter factor is the coefficient multiplying the octet component in the wave function of $\eta$ ($\eta^\prime$) in terms of the singlet and octet of SU(3)~\cite{Oset:2010tof}.

Considering the amplitudes of Eqs.~(\ref{tRP}) and (\ref{tRV}), we can determine the sum over the polarizations of 
\begin{align}
|t_{P_s\to P^\prime R}|^2=|t^\text{pseudo}_{P_s\to P^\prime R}+t^\text{vector}_{P_s\to P^\prime R}|^2,
\end{align}
finding
\begin{align}
&\sum\limits_\text{pol.}|t_{P_s\to P^\prime R}|^2=\sum\limits_\text{pol.}|t^\text{pseudo}_{P_s\to P^\prime R}|^2+\sum\limits_\text{pol.}|t^\text{vector}_{P_s\to P^\prime R}|^2\nonumber\\
&\quad+2\text{Re}\bigg\{\sum\limits_\text{pol.}t^\text{pseudo}_{P_s\to P^\prime R}\bigg(t^\text{vector}_{P_s\to P^\prime R}\bigg)^\dagger\bigg\},
\end{align}
where
\begin{align}
&\sum\limits_\text{pol.}|t^\text{pseudo}_{P_s\to P^\prime R}|^2=\frac{|g|^2}{4m_R m_{P_s}}\text{Tr}\Bigg\{(\slashed{P}-\slashed{k}+m_R)t^{(C)}_{\mu}(\pmb{C})\nonumber\\
&\quad\times (\slashed{P}+m_{P_s}) P^{\mu\sigma}t^{(C)}_{\sigma}(\pmb{C}^*)\Bigg\}=\frac{|g|^2}{4m_R m_{P_s}}\sum\limits_{l=0}^4 L^{(l)}(P\cdot k)^l,\nonumber\\
&\sum\limits_\text{pol.}|t^\text{vector}_{P_s\to P^\prime R}|^2=\frac{|G|^2}{24 m_Rm_{P_s}}\text{Tr}\Bigg\{(\slashed{p}+m_R)t^{(D)}_{\nu^\prime}(\pmb{D})(\slashed{P}+m_{P_s})\nonumber\\
&\quad\times P^{\nu^\prime\tilde{\nu}}t^{(D)}_{\tilde{\nu}}(\pmb{D}^*)\Bigg\}=\frac{|G|^2}{24 m_R m_{P_s}}\sum\limits_{l=0}^4M^{(l)}(P\cdot k)^l,\nonumber
\end{align}
\begin{align}
&\sum\limits_\text{pol.}t^\text{pseudo}_{P_s\to P^\prime R}\Bigg(t^\text{vector}_{P_s\to P^\prime R}\Bigg)^\dagger=\frac{g G}{4\sqrt{2}m_R m_{P_s}}\nonumber\\
&\quad\times\text{Tr}\Bigg\{(\slashed{p}+m_R)t^{(C)}_\mu(\pmb{C})(\slashed{P}+m_{P_s})P^{\mu\nu^\prime}t^{(D)}_{\nu^\prime}(\pmb{D}^*)\Bigg\}\nonumber\\
&\quad=i\frac{g G}{4\sqrt{2}m_Rm_{P_s}}\sum\limits_{l=0}^4 N^{(l)}(P\cdot k)^l,\label{Tr2}
\end{align}
with the coefficients $L^{(l)}$, $M^{(l)}$, and $N^{(l)}$ listed in the Appendix~\ref{FHJ}.

With the above amplitudes, the partial decay width of $P_s\to P^\prime_i B^\prime_i$, or $P_s\to P^\prime_i R_i$, can be determined from Eq.~(\ref{width}) replacing $\sum\limits_\text{pol.}|t_{P_s\to V_i B_i}|^2$ by either $\sum\limits_\text{pol.}|t_{P_s\to P^\prime_i B^\prime_i}|^2$ or $\sum\limits_\text{pol.}|t_{P_s\to P^\prime_i R_i}|^2$, and $m_{V_i}$  by $m_{P^\prime_i}$,  $m_{B_i}$ by $m_{R_i}$. The unstable character of the vector mesons in the intermediate states has been taken into account replacing $\omega_{V\,(V^\prime)}-i\epsilon$, with $\omega_{V\,(V^\prime)}$ representing their energies,  by $\omega_{V\,(V^\prime)}-i\Gamma_{V(V^\prime)}/2$, with $\Gamma_{V(V^\prime)}$ being their widths. In the case of having a resonance in the final state, its unstable character has been implemented by convoluting Eq.~(\ref{width}) with the corresponding spectral function for the resonance, i.e.,
\begin{align}
&\Gamma_{P_s\to P^\prime_i R_i}=\frac{1}{N_{P_s}N_{R_i}}\int\limits_{m_{P_s}-2\Gamma_{P_s}}^{m_{P_s}+2\Gamma_{P_s}} d\tilde{m}_{P_s}\rho_{P_s}(\tilde{m}_{P_s})\nonumber\\
&\quad\times\int\limits_{m_{R_i}-2\Gamma_{R_i}}^{m_{R_i}+2\Gamma_{R_i}} d\tilde{m}_{R_i}\rho_{R_i}(\tilde{m}_{R_i})\Gamma_{P_s\to P^\prime_i R_i}(\tilde{m}_{P_s},m_{P^\prime_i},\tilde{m}_{R_i}),\nonumber
\end{align}
with 
\begin{align}
\rho_{R_i}(\tilde{m}_{R_i})&=-\frac{1}{\pi}\text{Im}\Bigg(\frac{1}{\tilde{m}_{R_i}-m_{R_i}+i\Gamma_{R_i}/2}\Bigg),\nonumber\\
N_{R_i}&=\int\limits_{m_{R_i}-2\Gamma_{R_i}}^{m_{R_i}+2\Gamma_{R_i}}d\tilde{m}_{R_i}\rho_{R_i}(\tilde{m}_{R_i}).
\end{align}

\section{Results}
In Table~\ref{RNC} we show the partial decay widths obtained for the processes $P_s(2080)\to P^\prime B$, $P^\prime R$ without considering the convolution over the widths of the $P_s$ and of the resonance $R$ in the final state. 
\begin{table*}[t]
\caption{Partial decay widths (in MeV) of $P_s(2080)$ to final states formed by a pseudoscalar and an octet baryon and a pseudoscalar and $\Lambda(1405)/N^*(1535)/N^*(1650)$. We present the results obtained by considering the triangular loop mechanism of Fig.~\ref{Psdecay} including only the exchange of pseudoscalars between the vector and baryon produced from the primary vertex (P exch.) and considering the exchange of vector mesons too (P+V exch.). Here $\Lambda_1(1405)$ [$\Lambda_2(1405)$] represents the lower (higher) pole related to $\Lambda(1405)$~\cite{Oset:1997it,Jido:2003cb,Khemchandani:2018amu}.}\label{RNC}
\begin{tabular}{cccccc}
\text{channel}&\multicolumn{2}{c}{\text{width}}&\text{channel}&\multicolumn{2}{c}{\text{width}}\\
\hline
& P exch.& P+V exch.& &P exch. &P+V exch.\\
\hline\hline
$\pi^+ n$&$0.77\pm 0.21$&$0.95\pm 0.27$&$K^+\Lambda_2(1405)$&$5.05\pm0.76$&$5.10\pm0.77$\\
$\pi^0 p$&$0.38\pm 0.11$&$0.47\pm 0.13$&$\pi^+ N^{*0}(1535)$&$1.18\pm0.28$&$1.18\pm0.28$\\
$\eta p$&$0.87\pm 0.23$&$0.77\pm 0.21$&$\pi^0 N^{*+}(1535)$&$0.59\pm0.14$&$0.59\pm0.14$\\
$K^+\Lambda$&$3.83\pm 0.84$&$3.74\pm 0.82$&$\eta N^{*0}(1535)$&$0.33\pm0.05$&$0.34\pm0.06$\\
$K^+ \Sigma^0$&$1.56\pm 0.31$&$1.45\pm 0.29$&$\pi^+ N^{*0}(1650)$&$0.34\pm0.03$&$0.26\pm0.02$\\
$K^0 \Sigma^+$&$3.11\pm 0.62$&$2.90\pm 0.57$&$\pi^0 N^{*+}(1650)$&$0.17\pm0.01$&$0.13\pm0.01$\\
$\eta^\prime p$&$0.014\pm 0.005$&$0.07\pm 0.02$& & & \\
$K^+ \Lambda_1(1405)$&$16.97\pm2.67$&$17.16\pm2.71$& & \\
\end{tabular}
\end{table*}
The central values obtained represent the average values resulting from consideration of different form factors at the vertices, different cut-offs in those form factors as well as different $\eta-\eta^\prime$ mixing angle, while the uncertainty shown in the results of Table~\ref{RNC} corresponds to the standard deviation obtained. As can be seen, the contribution to the partial decay widths of diagrams in which a vector is exchanged between the vector and baryon produced at the primary vertex of the diagram in Fig.~\ref{Psdecay} is small, except for the case of the $\eta^\prime p$ final state.

It is interesting to notice that the partial decay width of $P_s(2080)$ to pseudoscalar-baryon channels with hidden strangeness, like $K\Lambda$ ($\Gamma_{K\Lambda}=3.74\pm 0.82$ MeV) and $K\Sigma$ ($\Gamma_{K\Sigma}=\Gamma_{K^+\Sigma^0}+\Gamma_{K^0\Sigma^+}=4.35\pm0.86$), is larger than the partial decay width to a channel like $\pi N$ ($\Gamma_{\pi N}=\Gamma_{\pi^+n}+\Gamma_{\pi^0p}=1.42\pm 0.40$ MeV). This result suggests that considering reactions in which $P_s(2080)$ is produced and decays to a final state like $K\Sigma$ and $K\Lambda$ can be more relevant than those involving $\pi N$ for identifying the generation of $P_s(2080)$. But it is even more interesting the fact that the partial decay width of $P_s(2080)$ to a final state formed by $\pi N^*(1535)$, for which $\Gamma_{\pi N^*(1535)}=\Gamma_{\pi^+ N^{*0}(1535)}+\Gamma_{\pi^0 N^{*+}(1535)}=1.77\pm 0.42$ MeV, is also comparable to the previous partial decay widths. There are several studies suggesting that $N^*(1535)$ has a sizeable hidden strangeness $K\Sigma$ component in its wave function~\cite{Inoue:2001ip,Nieves:2001wt,Khemchandani:2013nma}, producing a partial decay width of $P_s(2080)$ which is similar to that of $\pi N$, even if there is more phase space available for the latter channel. 

As can be seen in Table~\ref{RNC}, the decay of $P_s(2080)$ to $K\Lambda_1(1405)$ produces the largest contribution of the final states considered. Here we denote as $\Lambda_1(1405)$ and $\Lambda_2(1405)$ to the lower and upper mass poles, respectively, obtained in Refs.~\cite{Oset:1997it,Jido:2003cb,Khemchandani:2018amu} , where a double pole structure is suggested for $\Lambda(1405)$, with the lower (upper) pole having a mass $\sim 1380$ ($1426$) MeV and a larger coupling to the $\pi\Sigma$ ($\bar K N$) channel. In this way, reactions with a final state like $K\pi\Sigma$, where $\pi \Sigma$ has its origin in the decay of $\Lambda(1405)$, can be very relevant to extract information on the properties of $P_s(2080)$.

Considering all the partial decay widths listed in Table~\ref{RNC}, we obtain a width of $\sim 35\pm 6$ MeV, which is to be added to the width of $\sim 60-70$ MeV obtained from vector-baryon channels in Ref.~\cite{Khemchandani:2013nma}. Using as an estimation for the total width of $P_s(2080)$ a value of $\sim 100$ MeV, we can determine the partial decay widths of $P_s$ to vector-baryon channels. We can also estimate the effect of convoluting the partial decay widths of $P_s(2080)$ to $P^\prime B^\prime$ and $P^\prime R$ with the spectral function related to $P_s$ and, for the $P^\prime R$ channels, we can incorporate the finite width of the resonances $R$ in the calculation of the partial decay widths of $P_s(2080)$. The results obtained are similar to those found without implementing such effects, with the exception that when varying the masses of $P_s$ and $R$, the channel $\eta N^{*+}(1650)$ would be open for decay, finding a very small partial decay width ($\sim 0.005$ MeV). 

In Table~\ref{VB} we list the partial decay widths of $P_s(2080)$ to the vector-baryon channels considered in Refs.~\cite{Khemchandani:2011mf,Khemchandani:2013nma}. As can be seen, the largest contribution to the width comes from the $K^*\Sigma$ channel, to which $P_s$ couples more strongly, and whose nominal threshold is slightly above the mass of $P_s$, thus, the convolution here plays a relevant role for the calculation of the decay widths.

\begin{table}
\caption{Partial decay widths (in MeV) of $P_s(2080)$ to the vector-baryon channels (in the isospin $1/2$ basis) used for its generation in Refs.~\cite{Khemchandani:2011mf,Khemchandani:2013nma}.}\label{VB}
\begin{tabular}{cc}
\text{channel}&\text{width}\\
\hline\hline
$\rho N$& $5.66$\\
$\omega N$& $1.33$\\
$\phi N$& $1.92$\\
$K^*\Lambda$& $6.64$\\
$K^*\Sigma$& $49.97$
\end{tabular}
\end{table}

It should also be mentioned that the consideration of all the VB channels listed in Table~\ref{VB} is necessary when determining the partial decay widths of $P_s\to P^\prime B^\prime$, $P^\prime R$ via the triangular loop mechanism shown in Fig.~\ref{Psdecay}. For instance, considering only the primary vertex $P_s\to K^*\Sigma$, in view that the coupling constant of $P_s$ to $K^*\Sigma$ is the largest, drastically reduces the partial decay widths found. For example, to mention a few cases, the partial decay width to $\pi N$ would be $\sim 26$ times smaller, to $K\Sigma$ it will be a factor of $\sim 2$ smaller, and to $\eta N$ about 3 times smaller. 

\section{Conclusions}
The interest in studying the existence of $N^*$ resonances with hidden strangeness and masses around 2000 MeV has grown since the discovery of the $P_c$ states by the LHCb collaboration. Understanding the existence of lighter partners of these $P_c$ states with hidden strange content is part of the program of several experimental collaborations. However, detecting such states can be challenging due to the existence of several $N^*$ resonances in the same energy region. For this reason, studying the decay properties of these states and proposing non-standard final states, where the hidden strange quark content of the state could play a major role, is important for a better understanding of the properties of these states. In this work, we have focussed our attention on the $J^P=3/2^-$ $P_s(2080)$ state generated from vector-baryon dynamics in Refs.~\cite{Khemchandani:2011mf,Khemchandani:2013nma} and show that its partial decay widths to channels like $K\Sigma$, $K\Lambda$, $\pi N^*(1535)$ are as big as that to $\pi N$, with the decay to $K\Lambda(1405)$ giving a larger contribution. In this way, considering final states like $K\pi\Sigma$ could be relevant to understanding the properties of $P_s(2080)$.

\section{Acknowledgements}
This study was partly financed by the Coordena\c c\~ao de Aperfei\c coamento de Pessoal de N\'ivel Superior $-$ Brasil (CAPES) $-$ Finance Code 001. The partial support from other Brazilian agencies is also gratefully acknowledged: We thank CNPq (K.P.K: Grants No. 407437/ 2023-1 and No. 306461/2023-4; A.M.T: Grant No. 304510/2023-8), FAPESP (K.P.K.: Grant Number 2022/08347-9; A. M. T.: Grant number 2023/01182-7). The work of S.i.N. is supported by a grant from the National Research Foundation of Korea (NRF), funded by the Korean government (MSIT) (NRF- 2022K2A9A1A06091761).
\appendix
\section{Evaluation of the tensor integrals appearing in the formalism}\label{TI}
To calculate the integrals in Eq.~(\ref{Itensor}), we use the Passarino-Veltman decomposition of tensor integrals~\cite{Passarino:1978jh}: Let us consider, for example, the tensor integral $I^{(1)}_{\nu\mu}$. As a consequence of the Lorentz covariance, we see from Eq.~(\ref{Itensor}) that $\mathbb{I}^{(1)}_{\mu\nu}$ can be written as a linear combination of the metric tensor $g_{\nu\mu}$ and combinations of the four-momenta $P$ and $k$ forming a symmetric tensor of rank 2 under the interchange of $\mu$ and $\nu$, i.e.,
\begin{align}
\mathbb{I}^{(1)}_{\nu\mu}&=a^{(1)}_1 g_{\nu\mu}+a^{(1)}_2 P_\nu P_\mu+a^{(1)}_3 k_\nu k_\mu\nonumber\\
&\quad+a^{(1)}_4(P_\nu k_\mu+P_\mu k_\nu).\label{I1}
\end{align}
The coefficients $a^{(1)}_i$, $i=1,2,\dots,4$, in Eq.~(\ref{I1}) can be determined by contracting the latter equation by the different Lorentz structures present on it, i.e., $g^{\nu\mu}$, $P^\nu P^\mu$, $k^\nu k^\mu$ and $P^\nu k^\mu$. In this way, we can form a system of four coupled equations that allow us to express the coefficients $a^{(1)}_i$ in terms of the integrals 
\begin{align}
\mathbb{GI}^{(1)}&\equiv g^{\nu \mu}\mathbb{I}^{(1)}_{\nu\mu}=\int\limits_{-\infty}^{+\infty}\frac{d^4q}{(2\pi)^4}\frac{q^2}{\mathbb{D}},\nonumber\\
\mathbb{PPI}^{(1)}&\equiv P^\nu P^\mu \mathbb{I}^{(1)}_{\nu\mu}=\int\limits_{-\infty}^{+\infty}\frac{d^4q}{(2\pi)^4}\frac{(P\cdot q)^2}{\mathbb{D}},\nonumber\\
\mathbb{KKI}^{(1)}&\equiv k^\nu k^\mu \mathbb{I}^{(1)}=\int\limits_{-\infty}^{+\infty}\frac{d^4q}{(2\pi)^4}\frac{(k\cdot q)^2}{\mathbb{D}},\nonumber\\
\mathbb{PKI}^{(1)}&\equiv P^\nu k^\mu\mathbb{I}^{(1)}_{\nu\mu}=\int\limits_{-\infty}^{+\infty}\frac{d^4q}{(2\pi)^4}\frac{(P\cdot q)(k\cdot q)}{\mathbb{D}},\label{si}
\end{align}
as
\begin{align}
a^{(1)}_1&=-\frac{1}{2[(P\cdot k)^2-P^2 k^2]}\Bigg[\mathbb{GI}^{(1)}\{-(P\cdot k)^2+P^2k^2\}\nonumber\\
&\quad+2\mathbb{PKI}^{(1)} (P\cdot k)-\mathbb{PPI}^{(1)} k^2-\mathbb{KKI}^{(1)} P^2\Bigg],\nonumber
\end{align}
\begin{align}
a^{(1)}_2&=-\frac{1}{2[(P\cdot k)^2-P^2 k^2]^2}\Bigg[\mathbb{GI}^{(1)}k^2\{-(P\cdot k)^2+P^2k^2\}\nonumber\\
&\quad-\mathbb{KKI}^{(1)}\{ 2(P\cdot k)^2+P^2k^2\}\nonumber\\
&\quad+6\mathbb{PKI}^{(1)} k^2 (P\cdot k)-3\mathbb{PPI}^{(1)} k^4\Bigg],\nonumber
\end{align}
\begin{align}
a^{(1)}_3&=-\frac{1}{2[(P\cdot k)^2-P^2 k^2]^2}\Bigg[\mathbb{GI}^{(1)} P^2\{-(P\cdot k)^2+P^2 k^2\}\nonumber\\
&\quad-\mathbb{PPI}^{(1)}\{2(P\cdot k)^2+P^2 k^2\}\nonumber\\
&\quad+6\mathbb{PKI}^{(1)} P^2 (P\cdot k)-3\mathbb{KKI}^{(1)}P^4\Bigg],\nonumber
\end{align}
\begin{align}
a^{(1)}_4&=-\frac{1}{2[(P\cdot k)^2-P^2 k^2]^2}\Bigg[\mathbb{GI}^{(1)}(P\cdot k)\{(P\cdot k)^2-P^2 k^2\}\nonumber\\
&\quad-2\mathbb{PKI}^{(1)}\{2(P\cdot k)^2+P^2 k^2\}\nonumber\\
&\quad+3\mathbb{KKI}^{(1)}P^2(P\cdot k)+3\mathbb{PPI}^{(1)}k^2(P\cdot k)\Bigg].\label{a1i}
\end{align}

Next, to determine the integrals in Eq.~(\ref{si}), and find, in this way, the value of $a^{(1)}_i$, we first perform the $dq^0$ integration by considering Cauchy's theorem, finding
\begin{align}
\mathcal{I}_n=\int\limits_{-\infty}^{+\infty}\frac{dq^0}{2\pi}\frac{(q^0)^n}{\mathbb{D}}=-i\frac{N_n}{\mathcal{D}},\label{Nn}
\end{align}
where
\begin{align}
&\mathcal{D}=2\,\omega_B\,\omega_V\,\omega_P(P^0+\omega_B+\omega_V)(k^0+\omega_V+\omega_P)\nonumber\\
&\quad\times[P^0-k^0-\omega_B-\omega_P+i\epsilon][k^0-P^0-\omega_B-\omega_P+i\epsilon]\nonumber\\
&\quad\times[P^0-\omega_B-\omega_V+i\epsilon][k^0-\omega_V-\omega_P+i\epsilon],
\end{align}
with 
\begin{align}
\omega_B&=\sqrt{(\pmb{P}-\pmb{k}-\pmb{q})^2+m^2_B},\nonumber\\
\omega_V&=\sqrt{(\pmb{k}+\pmb{q})^2+m^2_V},\nonumber\\
\omega_P&=\sqrt{\pmb{q}^2+m^2_P}.
\end{align}
The numerators $N_n$ in Eq.~(\ref{Nn}) for the cases concerned in Eq.~(\ref{si}) are:
\begin{align}
&N_0=2k^0 P^0\omega_B\,\omega_P-(k^0)^2\omega_P\,\omega_{B+V}+\omega_{P+V}\nonumber\\
&\times[-(P^{0})^2\omega_B+\omega_{B+P}\,\omega_{B+V}\,\omega_{B+V+P}],\nonumber\\\nonumber\\
 &N_1=\omega_P\Big(-k^0 (P^0)^2\omega_B-(k^0)^3\omega_{B+V}+(k^0)^2P^0\nonumber\\
 &\times(2\omega_B+\omega_V)-P^0\omega_V\omega_{P+V}(2\omega_B+\omega_{P+V})\nonumber\\
 &+k^0\omega_{B+V}\big[\omega^2_B+\omega^2_{P+V}+\omega_B(2\omega_P+\omega_V)\big]\Big),\nonumber\\\nonumber\\
&N_2=\omega_P\Big((k^0)^2\omega_{B+V}\big[(P^0-k^0)^2-\omega^2_{B+P}-2\,\omega_P\,\omega_V\nonumber\\
&-\omega^2_V\big]+\omega_V\,\omega_{P+V}\big[\omega_B\,\omega_{B+P}\,\omega_{B+V}-(P^0)^2\omega_{B+P+V}\big]\nonumber\\
&+2k^0 P^0\omega_V[\omega^2_{P+V}+\omega_B(2\omega_P+\omega_V)]\Big),\nonumber
\end{align}
\begin{align}
&N_3=\omega_P\Big(-(k^0)^5\omega_{B+V}+(k^0)^4 P^0 (2\omega_B+3\omega_V)\nonumber\\
&+(k^0)^2 P^0\omega_V\big[(P^0)^2-\omega^2_B-3\omega^2_{P+V}-2\omega_B\nonumber\\
&\times(3\omega_P+\omega_V)\big]+(k^0)^3\big[-(P^0)^2(\omega_B+3\omega_V)\nonumber\\
&+\omega_{B+V}(\omega^2_{B+P}+2\omega_P\omega_V+\omega^2_V)\big]+P^0\omega_V\omega_{P+V}\nonumber\\
&\times[-(P^0)^2\omega_{P+V}+\omega_B(2\omega_P\omega_V+\omega_B\omega_{P+V})]\nonumber\\
&+k^0\omega_V\big[(P^0)^2\big(3\omega^2_{P+V}+\omega_B(2\omega_P+\omega_V)\big)\nonumber\\
&-\omega_B\omega_{B+V}\big(\omega_B(2\omega_P+\omega_V)+\omega_P(3\omega_P+2\omega_V)\big)\big]\Big),\nonumber\\\nonumber\\
&N_4=\omega_P\Big((k^0)^6\omega_{B+V}-2(k^0)^5 P^0(\omega_B+2\omega_V)+4 (k^0)^3 \nonumber\\
&\times P^0\omega_V\big[-(P^0)^2+\omega^2_{B+P}+(\omega_B+2\omega_P)\omega_V+\omega^2_V\big]\nonumber\\
&+(k^0)^4\big[(P^0)^2(\omega_B+6\omega_V)-\omega_{B+V}\big(\omega^2_{B+P}\nonumber\\
&+(\omega_B+2\omega_P)\omega_V+\omega^2_V\big)\big]-\omega_V\omega_{P+V}\big[(P^0)^4\omega_{P+V}\nonumber\\
&+\omega_B\omega_{B+P}\omega_{B+V}\big(\omega_P\omega_V+\omega_B\omega_{P+V}\big)-(P^0)^2\omega_B\nonumber\\
&\times\big(\omega^2_P+3\omega_P\omega_V+\omega^2_V+2\omega_B\omega_{P+V}\big)\big]+2k^0 P^0\omega_V\nonumber\\
&\times\big[2(P^0)^2\omega^2_{P+V}-\omega_B\big(2\omega_B\omega^2_{P+V}+\omega_V(2\omega_P+\omega_V)^2\big)\big]\nonumber\\
&+(k^0)^2\omega_V\big[(P^0)^4-2(P^0)^2\big(\omega^2_B+3\omega^2_{P+V}+\omega_B(2\omega_P+\omega_V)\big)\nonumber\\
&+\omega_B\omega_{B+V}\big(\omega^2_B+6\omega^2_P+4\omega_P\omega_V+\omega^2_V+\omega_B(4\omega_P+\omega_V)\big)\big]\Big),\nonumber
\end{align}

\begin{align}
&N_5=\omega_P\Big(-(k^0)^7\omega_{B+V}+(k^0)^6P^0(2\omega_B+5\omega_V)\nonumber\\
&-\omega_V\omega_{P+V}\big[(P^0+\omega_B)(P^0)^4\omega_{P+V}+\omega_B\omega^2_{B+P}\omega^2_{B+V}\omega_{P+V}\nonumber\\
&-2(P^0)^3\omega_B\omega_{P+V}\omega_{B+P+V}+P^0\omega_B\big(\omega^2_B\omega_P(\omega_B+2\omega_P)\nonumber\\
&+\omega_B(\omega_B+2\omega_P)^2\omega_V+2\omega^2_{B+P}\omega^2_V\big)-(P^0)^2\omega_B\omega_{P+V}\nonumber\\
&\times\big(2\omega^2_B+\omega^2_P+\omega^2_V+2\omega_B\omega_{P+V}\big)\big]-(k^0)^4\omega_V\nonumber\\
&\times\big[-10(P^0)^3-(P^0)^2\omega_B+\omega_B\omega^2_{B+V}+5 P^0\nonumber\\
&\times\big(2\omega^2_B+2\omega_B\omega_{P+V}+\omega^2_{P+V}\big)\big]-(k^0)^3\omega_V\big[5(P^0)^4\nonumber\\
&+2(P^0)^3\omega_B-2P^0\omega_B\omega^2_{B+V}-2(P^0)^2\big(4\omega^2_B+4\omega_B\omega_{P+V}\nonumber\\
&+5\omega^2_{P+V}\big)+\omega_B\omega_{B+V}\big(3\omega^2_B+10\omega^2_P+8\omega_P\omega_V+3\omega^2_V\nonumber\\
&+\omega_B(8\omega_P+\omega_V)\big)\big]+(k^0)^5\big[-(P^0)^2(\omega_B+10\omega_V)\nonumber\\
&+\omega_{B+V}\big(\omega^2_B+\omega^2_{P+V}+\omega_B(2\omega_P+3\omega_V)\big)\big]\nonumber\\
&+k^0\omega_V\big[5(P^0)^4\omega^2_{P+V}+2(P^0)^3\omega_B\omega^2_{P+V}\nonumber\\
&-2 P^0\omega_B\omega^2_{B+V}\omega^2_{P+V}-(P^0)^2\omega_B\big(4\omega^3_P+18\omega^2_P\omega_V\nonumber\\
&+20\omega_P\omega^2_V+7\omega^3_V+8\omega_B\omega^2_{P+V}\big)+\omega_B\omega_{B+V}\nonumber\\
&\times\big(3\omega^2_B\omega^2_{P+V}+\omega^2_P\omega_V(4\omega_P+3\omega_V)+\omega_B\omega_P\nonumber\\
&\times(\omega_P+2\omega_V)(4\omega_P+3\omega_V)\big)\big]+(k^0)^2\omega_V\nonumber\\
&\times\big[(P^0)^5+(P^0)^4\omega_B+\omega_B\omega^2_{B+V}\big(\omega^2_{B+P}+\omega^2_{P+V}\big)\nonumber\\
&-2(P^0)^2\omega_B\big(\omega^2_B+\omega^2_P+\omega_P\omega_V+\omega^2_V+\omega_B\omega_{P+V}\big)\nonumber\\
&-2(P^0)^3\big(\omega^2_B+\omega_B\omega_{P+V}+5\omega^2_{P+V}\big)+P^0\omega_B\nonumber\\
&\times\big(\omega^3_B+2\omega^2_B\omega_{P+V}+2\omega_V\big(10\omega^2_P+11\omega_P\omega_V\nonumber\\
&+4\omega^2_V\big)+2\omega_B\big(5\omega^2_P+12\omega_P\omega_V+5\omega^2_V\big)\big)\big]\Big),
\end{align}
where 
\begin{align}
&\omega_{B+V}=\omega_B+\omega_V,\nonumber\\
&\omega_{B+P}=\omega_B+\omega_P,\nonumber\\
&\omega_{P+V}=\omega_P+\omega_V,\nonumber\\
&\omega_{B+P+V}=\omega_B+\omega_P+\omega_V.
\end{align}
Here we consider the rest frame of the decaying particle to determine the partial decay widths of $P_s(2080)$, thus, $\pmb{P}=\pmb{0}$, $P^0=\sqrt{s}=m_{P_s}$ and
\begin{align}
k^0=\frac{s+m^2_{P^\prime}-m^2_{B^\prime\,(R)}}{2\sqrt{s}},
\end{align}
with $m_{P^\prime}$, $m_{B^\prime\,(R)}$ being, respectively, the masses of the pseudoscalar and baryon (resonance) in the final state.

Using Eq.~(\ref{Nn}), we can write the integrals in Eq.~(\ref{si}) as
\begin{align}
\mathbb{GI}^{(1)}&=\int\limits_{-\infty}^{+\infty}\frac{d^3q}{(2\pi)^3}[\mathcal{I}_2-\mathcal{I}_0\pmb{q}^2],\nonumber\\
\mathbb{PPI}^{(1)}&=\int\limits_{-\infty}^{+\infty}\frac{d^3q}{(2\pi)^3}(P^0)^2\mathcal{I}_2,\nonumber\\
\mathbb{KKI}^{(1)}&=\int\limits_{-\infty}^{+\infty}\frac{d^3q}{(2\pi)^3}[(k^0)^2\mathcal{I}_2-2k^0(\pmb{k}\cdot\pmb{q})\mathcal{I}_1+(\pmb{k}\cdot\pmb{q})^2\mathcal{I}_0],\nonumber\\
\mathbb{PKI}^{(1)}&=\int\limits_{-\infty}^{+\infty}\frac{d^3q}{(2\pi)^3}[P^0 k^0\mathcal{I}_2-P^0(\pmb{k}\cdot\pmb{q})\mathcal{I}_1],
\end{align}
and the $d^3q$ integral is given by
\begin{align}
\int\limits_{-\infty}^{+\infty}\frac{d^3 q}{(2\pi)^3}=\int\limits_0^\infty d|\pmb{q}|\int\limits_{-1}^1d\text{cos}\theta |\pmb{q}|^2\frac{1}{(2\pi)^2}.
\end{align}
The $d|\pmb{q}|$ integral is regularized with form factors. We consider either Gaussian ($F_G$), Lorentzian ($F_L$), or Heavise ($F_H$) theta-function form factors at each vertex, i.e.,
\begin{align}
F_G(\pmb{q})&=e^{-|\pmb{q}|^2/(2\Lambda^2)},\nonumber\\
F_L(\pmb{q})&=\frac{\Lambda^2}{\Lambda^2+|\pmb{q}|^2},\nonumber\\
F_H(\pmb{q})&=\Theta(\Lambda-|\pmb{q}|),
\end{align}
where $\Lambda\sim 600-900$ MeV. To compare results obtained with different form factors, we consider the normalization 
\begin{align}
\int\limits_0^\infty d|\pmb{q}|F^2_H(\pmb{q})=\int\limits_0^\infty d|\pmb{q}|F^2_G(\pmb{q})=\int\limits_0^\infty d|\pmb{q}|F^2_L(\pmb{q}),
\end{align}
which implies a different value of $\Lambda$ for each type of form factor. When considering final states involving a resonance, the cut-off $\Lambda$ used is for the modulus of the center of mass momentum of the particles forming the resonance, thus, a boost needs to be performed from the rest frame of the decaying particle to the rest frame of the resonance in the final state.

Similarly, we can write
\begin{align}
\mathbb{I}^{(2)}_{\nu\mu}&=a^{(2)}_1 g_{\nu\mu}+a^{(2)}_2 P_\nu P_\mu+a^{(2)}_3 k_\nu k_\mu\nonumber\\
&\quad+a^{(2)}_4(P_\nu k_\mu+P_\mu k_\nu),\label{I2}
\end{align}
and
\begin{align}
\mathbb{I}^{(i)}_\nu&=a^{(i)}_1P_\nu+a^{(i)}_2k_\nu,
\end{align}
with $i=3,4,5$, where the coefficients $a^{(2)}_i$ can be obtained from Eq.~(\ref{a1i}) by changing $\mathbb{GI}^{(1)}$, $\mathbb{PPI}^{(1)}$, $\mathbb{KKI}^{(1)}$ and $\mathbb{PKI}^{(1)}$ to
\begin{align}
\mathbb{GI}^{(2)}&=\int\limits_{-\infty}^{+\infty}\frac{d^4q}{(2\pi)^4}\frac{q^4}{\mathbb{D}}=\int\limits_{-\infty}^{+\infty}\frac{d^3q}{(2\pi)^3}(\mathcal{I}_4-2\pmb{q}^2\mathcal{I}_2+\pmb{q}^4\mathcal{I}_0),\nonumber\\
\mathbb{PPI}^{(2)}&=\int\limits_{-\infty}^{+\infty}\frac{d^4q}{(2\pi)^4}\frac{q^2 (P\cdot q)^2}{\mathbb{D}}=\int\limits_{-\infty}^{+\infty}\frac{d^3q}{(2\pi)^3}(P^0)^2(\mathcal{I}_4-\pmb{q}^2\mathcal{I}_2),\nonumber\\
\mathbb{KKI}^{(2)}&=\int\limits_{-\infty}^{+\infty}\frac{d^4q}{(2\pi)^4}\frac{q^2(k\cdot q)^2}{\mathbb{D}}=\int\limits_{-\infty}^{+\infty}\frac{d^3q}{(2\pi)^3}[(k^0)^2\mathcal{I}_4\nonumber\\
&\quad-2k^0(\pmb{k}\cdot\pmb{q})\mathcal{I}_3+\big((\pmb{k}\cdot\pmb{q})^2-(k^0)^2\pmb{q}^2\big)\mathcal{I}_2\nonumber\\
&\quad+2k^0(\pmb{k}\cdot\pmb{q})\pmb{q}^2\mathcal{I}_1-\pmb{q}^2(\pmb{k}\cdot\pmb{q})^2\mathcal{I}_0],\nonumber\\
\mathbb{PKI}^{(2)}&=\int\limits_{-\infty}^{+\infty}\frac{d^4q}{(2\pi)^4}\frac{q^2(P\cdot q)(k\cdot q)}{\mathbb{D}}=\int\limits_{-\infty}^{+\infty}\frac{d^3q}{(2\pi)^3}P^0\nonumber\\
&\quad\times[k^0\mathcal{I}_4-\pmb{k}\cdot\pmb{q}\mathcal{I}_3-k^0\pmb{q}^2\mathcal{I}_2+\pmb{q}^2(\pmb{k}\cdot\pmb{q})\mathcal{I}_1],\nonumber
\end{align}
and
\begin{align}
a^{(i)}_1&=-\frac{k^2 \mathbb{PI}^{(i)}-(P\cdot k)\mathbb{KI}^{(i)}}{(P\cdot k)^2-k^2 P^2},\nonumber\\
a^{(i)}_2&=\frac{(P\cdot k) \mathbb{PI}^{(i)}-P^2\mathbb{KI}^{(i)}}{(P\cdot k)^2-k^2 P^2},
\end{align}
with $i=3,4,5$. 

Once we have determined the coefficients $a^{(i)}_j$, we define the following combinations appearing in Eq.~(\ref{Ai})
 \begin{align}
 A^{VBP}_1&=-\Bigg[\bigg(1+\frac{k^2}{m^2_V}\bigg)a^{(1)}_1-\frac{1}{m^2_V}a^{(2)}_1\Bigg],\nonumber\\
 A^{VBP}_2&=-\Bigg[\bigg(1+\frac{k^2}{m^2_V}\bigg)a^{(1)}_2-\frac{1}{m^2_V}a^{(2)}_2\Bigg],\nonumber\\
 A^{VBP}_3&=-a^{(1)}_3+a^{(3)}_2-\frac{k^2}{m^2_V}(a^{(1)}_3+a^{(3)}_2)\nonumber\\
 &\quad+\frac{1}{m^2_V}(a^{(2)}_3+a^{(4)}_2),\nonumber\\
 A^{VBP}_4&=-a^{(1)}_4+a^{(3)}_1-\frac{k^2}{m^2_V}(a^{(1)}_4+a^{(3)}_1)\nonumber\\
 &\quad+\frac{1}{m^2_V}(a^{(2)}_4+a^{(4)}_1),\nonumber\\
 A^{VBP}_5&= -\Bigg[\bigg(1+\frac{k^2}{m^2_V}\bigg)a^{(1)}_4-\frac{1}{m^2_V}a^{(2)}_4\Bigg],\nonumber\\
 \end{align}
 \begin{align}
  A^{VBP}_6&= \bigg(1+\frac{k^2}{m^2_V}\bigg)a^{(4)}_1-\frac{1}{m^2_V}a^{(5)}_1,\nonumber\\
 A^{VBP}_7&=a^{(4)}_2-4 a^{(1)}_1-a^{(1)}_2P^2-a^{(1)}_3k^2-2a^{(1)}_4 (P\cdot k)\nonumber\\
 &\quad+\frac{k^2}{m^2_V}\bigg(a^{(4)}_2+4a^{(1)}_1+a^{(1)}_2 P^2+a^{(1)}_3k^2\nonumber\\
 &\quad+2 a^{(1)}_4 (P\cdot k)\bigg)-\frac{1}{m^2_V}\bigg(a^{(5)}_2+4 a^{(2)}_1\nonumber\\
 &\quad+a^{(2)}_2 P^2+a^{(2)}_3 k^2+2 a^{(2)}_4 (P\cdot k)\bigg).
 \end{align}
 
 Next, when dealing with a particular vector-vector-baryon intermediate state, $VV^\prime B$, in the triangular loop, we define the coefficients
\begin{align}
&B^{VV^\prime B}_1=b^{(1)}_1;~B^{VV^\prime B}_2=b^{(1)}_2;~B^{VV^\prime B}_3=b^{(1)}_4;\nonumber\\
&B^{VV^\prime B}_4=b^{(3)}_1 m_B;~B^{VV^\prime B}_5=b^{(3)}_1,\label{Bcof1}
\end{align}
which appear in Eq.~(\ref{Bi}). Here, the coefficients $b^{(i)}_j$ are analogous to $a^{(i)}_j$ but replacing $\omega_P$ by $\omega_{V^\prime}$.

In case of Eq.~(\ref{Ci}), we have the following coefficients:
\begin{align}
&C^{VBP}_l=-A^{VBP}_l,~l=1,2,\dots,5\nonumber\\
&C^{VBP}_6=(m_R+m_B)\bigg[-\bigg(1+\frac{k^2}{m^2_V}\bigg)a^{(3)}_1+\frac{a^{(4)}_1}{m^2_V}\bigg],\nonumber\\
&C^{VBP}_7=(m_R+m_B)\bigg[\mathbb{I}^8-a^{(3)}_2-\frac{k^2}{m^2_V}(a^{(3)}_2+\mathbb{I}^{(8)})\nonumber\\
&\quad+\frac{1}{m^2_V}\bigg(a^{(4)}_2+4a^{(1)}_1+a^{(1)}_2P^2+a^{(1)}_3k^2+2a^{(1)}_4 P\cdot k\bigg)\bigg].\label{C7}
\end{align}
where the integral $\mathbb{I}^{(8)}$ is given by
\begin{align}
\mathbb{I}^{(8)}=\int\limits_{-\infty}^{+\infty}\frac{d^3q}{(2\pi)^3}\mathcal{I}_0.
\end{align}

To determine Eq.~(\ref{Di}), we need the coefficients $D^{VBV^\prime}_i$, which are given by
\begin{align}
&D^{VBV^\prime}_1=b^{(1)}_1;~ D^{VBV^\prime}_2=b^{(1)}_2, \nonumber\\
&D^{VBV^\prime}_3=b^{(1)}_4;~D^{VBV^\prime}_4=(m_R+m_B)b^{(3)}_1,
\end{align}
where $m_B\,(m_R)$ represents the mass of the baryon (resonance) in the intermediate (final) state.

\section{Coefficients $C_{PB\to B^\prime}$, $\dots$, $C_{V\to V^\prime P^\prime}$}\label{coeff}
In tables~\ref{T1}-\ref{Tf}, we provide the coefficients $C_{PB\to B^\prime}$, $C_{V\to P^\prime P}$, $C_{V^\prime B\to B^\prime}$, $C_{V\to V^\prime P^\prime}$ needed to evaluate the amplitudes associated with the triangular diagrams shown in Fig.~\ref{Psdecay} for the different final and intermediate states. In the case of the coefficients related to vector-pseudoscalar-pseudoscalar, baryon-baryon-pseudoscalar, vector-vector-pseudoscalar vertices, we have considered an $\eta-\eta^\prime$ mixing angle in the range $\beta\simeq-15^\circ$ to $-22^\circ$~\cite{Gilman:1987ax,Akhoury:1987ed,Bramon:1997mf} instead of assuming ideal mixing, which corresponds to an angle $\beta$ with $\text{sin}\beta=-1/3$, i.e., $\beta\simeq -19.43^\circ$. 

It should be noted that for a general mixing angle $\beta$, the matrix $P$ related to the pseudoscalar fields in Eq.~(\ref{PVB}) reads as~\cite{Malabarba:2023zez}
\begin{widetext}
\begin{align}
\mathbb{P}&=\left(\begin{array}{ccc}A(\beta)\eta+B(\beta)\eta^\prime+\frac{\pi^0}{\sqrt{2}}&\pi^+&K^+\\\pi^-&A(\beta)\eta+B(\beta)\eta^\prime-\frac{\pi^0}{\sqrt{2}}&K^0\\K^-&\bar K^0&C(\beta)\eta+D(\beta)\eta^\prime\end{array}\right),\label{Pmat}
\end{align}
\end{widetext}
where
\begin{align}
A(\beta)&=-\frac{\text{sin}\beta}{\sqrt{3}}+\frac{\text{cos}\beta}{\sqrt{6}},\nonumber\\
B(\beta)&=\frac{\text{sin}\beta}{\sqrt{6}}+\frac{\text{cos}\beta}{\sqrt{3}},\nonumber
\end{align}
\begin{align}
C(\beta)&=-\frac{\text{sin}\beta}{\sqrt{3}}-\sqrt{\frac{2}{3}}\text{cos}\beta,\nonumber\\
D(\beta)&=-\sqrt{\frac{2}{3}}\text{sin}\beta+\frac{\text{cos}\beta}{\sqrt{3}}.
\end{align}

\begin{table}[h!]
\caption{Coefficients $C_{V\to PP^\prime}$, $C_{BP\to B^\prime}$, $C_{V\to V^\prime P^\prime}$ and $C_{B V^\prime\to B^\prime}$ for the final state $K^0\Sigma^+$. To simplify the notation, we define $C_\beta=\text{cos}\beta$ and $S_\beta=\text{sin}\beta$.}\label{T1}
\begin{tabular}{ccc}
\hline\hline
$VBP$&$C_{V\to PP^\prime}$&$C_{B P\to B^\prime}$\\
\hline\\
$K^{*0}\Sigma^+\pi^0$&$-1/\sqrt{2}$&$-F/f_\pi$\\
$K^{*0}\Sigma^+\eta$&$\sqrt{3/2}C_\beta$&$(-DC_\beta +\sqrt{2}DS_\beta)/(\sqrt{3} f_\pi)$\\
$K^{*0}\Sigma^+\eta^\prime$&$\sqrt{3/2} S_\beta$&$-D(\sqrt{2}C_\beta+S_\beta)/(\sqrt{3}f_\pi)$\\
$K^{*0}\Sigma^0\pi^+$&$1$&$F/f_\pi$\\
$K^{*0}\Lambda\pi^+$&$1$&$-D/(\sqrt{3}f_\pi)$\\
$\rho^0 p\bar K^0$&$1/\sqrt{2}$&$(-D+F)/(\sqrt{2}f_\pi)$\\
$\omega p\bar K^0$&$-1/\sqrt{2}$&$(-D+F)/(\sqrt{2}f_\pi)$\\
$\phi p \bar K^0$& 1&$(-D+F)/(\sqrt{2}f_\pi)$\\\\
\hline
$VBV^\prime$&$C_{V\to V^\prime P^\prime}$&$C_{B V^\prime\to B^\prime}$\\
\hline\\
$K^{*0}\Sigma^+\rho^0$&$-1/\sqrt{2}$&$\sqrt{2}$\\
$K^{*0}\Sigma^+\omega$&$1/\sqrt{2}$&$\sqrt{2}$\\
$K^{*0}\Sigma^+\phi$&$1$&$1$\\
$K^{*0}\Sigma^0\rho^+$&$1$&$-\sqrt{2}$\\
$K^{*0}\Lambda\rho^+$&$1$&$0$\\
$\rho^0 p\bar K^{*0}$&$-1/\sqrt{2}$&$-1$\\
$\omega p\bar K^{*0}$&$1/\sqrt{2}$&$-1$\\
$\phi p\bar K^{*0}$&$1$&$-1$
\end{tabular}
\end{table}

\begin{table}[h!]
\caption{Same as in Table~\ref{T1} but for the $K^+\Sigma^0$, $K^+\Lambda$ and $K^+\Lambda(1405)$ final states. In the case of $P_s\to K^+\Lambda(1405)$, $C_{BP\to B^\prime}=g_{\Lambda^*\to PB}$, $C_{BV^\prime\to B^\prime}=g_{\Lambda^*\to V^\prime B}$, i.e., the coupling constants of the resonance to the PB and VB channels. Here, by $\Lambda(1405)$, we refer to any of the two poles obtained in Ref.~\cite{Jido:2003cb,Khemchandani:2011mf}.}
\resizebox{0.5\textwidth}{!}{
\begin{tabular}{cccc}
\hline\hline\\
$VBP$&$C_{V\to PP^\prime}$&\multicolumn{2}{c}{$C_{B P\to B^\prime}$}\\
&$K^+\Sigma^0/K^+\Lambda/K^+\Lambda(1405)$&$K^+\Sigma^0$&$K^+\Lambda$\\
\hline\\
$K^{*0}\Sigma^+\pi^-$&$1$&$F/f_\pi$&$-D/(\sqrt{3}f_\pi)$\\
$K^{*+}\Sigma^0\pi^0$&$1/\sqrt{2}$&0&$-D/(\sqrt{3}f_\pi)$\\
$K^{*+}\Sigma^0\eta$&$\sqrt{3/2} C_\beta$&$(-DC_\beta +\sqrt{2}DS_\beta)/(\sqrt{3} f_\pi)$&$0$\\
$K^{*+}\Sigma^0\eta^\prime$&$\sqrt{3/2}S_\beta$&$-D(\sqrt{2}C_\beta+S_\beta)/(\sqrt{3}f_\pi)$&$0$\\
$K^{*+}\Lambda\pi^0$&$1/\sqrt{2}$&$-D/(\sqrt{3}f_\pi)$&$0$\\
$K^{*+}\Lambda\eta$&$\sqrt{3/2} C_\beta$&$0$&$D(C_\beta+\sqrt{2}S_\beta)/(\sqrt{3}f_\pi)$\\
$K^{*+}\Lambda\eta^\prime$&$\sqrt{3/2} S_\beta$&$0$&$D(-\sqrt{2}C_\beta+S_\beta)/(\sqrt{3}f_\pi)$\\
$\rho^0 p K^-$&$-1/\sqrt{2}$&$(-D+F)/(2f_\pi)$&$(D+3F)/(2\sqrt{3}f_\pi)$\\
$\rho^+ n \bar K^0$&$-1$&$(D-F)/(2f_\pi)$&$(D+3F)/(2\sqrt{3}f_\pi)$\\
$\omega p K^-$&$-1/\sqrt{2}$&$(-D+F)/(2f_\pi)$&$(D+3F)/(2\sqrt{3}f_\pi)$\\
$\phi p K^-$& 1&$(-D+F)/(2f_\pi)$&$(D+3F)/(2\sqrt{3}f_\pi)$\\\\
\hline
$VBV^\prime$&$C_{V\to V^\prime P^\prime}$&\multicolumn{2}{c}{$C_{B V^\prime\to B^\prime}$}\\

&$K^+\Sigma^0/K^+\Lambda/K^+\Lambda(1405)$&$K^+\Sigma^0$&$K^+\Lambda$\\
\hline\\
$K^{*0}\Sigma^+\rho^-$&$1$&$-\sqrt{2}$&$0$\\
$K^{*+}\Sigma^0\rho^0$&$1/\sqrt{2}$&$0$&$0$\\
$K^{*+}\Sigma^0\omega$&$1/\sqrt{2}$&$\sqrt{2}$&$0$\\
$K^{*+}\Sigma^0\phi$&$1$&$1$&$0$\\
$K^{*+}\Lambda\rho^0$&$1/\sqrt{2}$&$0$&$0$\\
$K^{*+}\Lambda\omega$&$1/\sqrt{2}$&$0$&$\sqrt{2}$\\
$K^{*+}\Lambda\phi$&$1$&$0$&$1$\\
$\rho^0 p K^{*-}$&$1/\sqrt{2}$&$-1/\sqrt{2}$&$-\sqrt{3/2}$\\
$\rho^+ n \bar K^{*0}$&$1$&$1/\sqrt{2}$&$-\sqrt{3/2}$\\
$\omega p K^{*-}$&$1/\sqrt{2}$&$-1/\sqrt{2}$&$-\sqrt{3/2}$\\
$\phi p K^{*-}$& 1&$-1/\sqrt{2}$&$-\sqrt{3/2}$
\end{tabular}}
\end{table}

\begin{table}[h!]
\caption{Same as in Table~\ref{T1} but for the $\pi^0p$, and $\pi^0 N^{*+}$ final states. Here $N^{*+}$ represents either $N^{*+}(1535)$ or $N^{*+}(1650)$. In the case of $P_s\to \pi^0 N^{*+}$, $C_{BP\to B^\prime}=g_{N^{*+}\to PB}$, $C_{BV^\prime\to B^\prime}=g_{N^{*+}\to V^\prime B}$, i.e., the coupling constants of the resonance to the PB and VB channels.}
\resizebox{0.5\textwidth}{!}{
\begin{tabular}{ccc}
\hline\hline\\
$VBP$&$C_{V\to PP^\prime}$&$C_{B P\to B^\prime}$\\
&$\pi^0 p/\pi^0 N^{*+}$&$\pi^0 p$\\
\hline\\
$K^{*0}\Sigma^+ K^0$&$1/\sqrt{2}$&$(-D+F)/(\sqrt{2}f_\pi)$\\
$K^{*+}\Sigma^0 K^+$&$-1/\sqrt{2}$&$(-D+F)/(2f_\pi)$\\
$K^{*+}\Lambda K^+$&$-1/\sqrt{2}$&$(D+3F)/(2\sqrt{3}f_\pi)$\\
$\rho^0 p \pi^0$&$0$&$-(D+F)/(2f_\pi)$\\
$\rho^0 p \eta$&$0$&$(C_\beta(D-3F)+2\sqrt{2}DS_\beta)/(2\sqrt{3}f_\pi)$\\
$\rho^0 p \eta^\prime$&$0$&$(-2\sqrt{2}D C_\beta +(D-3F)S_\beta)/(2\sqrt{3}f_\pi)$\\
$\rho^+ n \pi^+$&$-\sqrt{2}$&$-(D+F)/(\sqrt{2}f_\pi)$\\
$\omega p \pi^0$&$0$&$-(D+F)/(2f_\pi)$\\
$\omega p \eta$&$0$&$(C_\beta (D-3F)+2\sqrt{2} DS_\beta)/(2\sqrt{3} f_\pi)$\\
$\omega p \eta^\prime$&$0$&$(-2\sqrt{2}D C_\beta +(D-3F)S_\beta)/(2\sqrt{3}f_\pi)$\\
$\phi p \pi^0$& 0&$-(D+F)/(2f_\pi)$\\
$\phi p \eta$& 0&$(C_\beta (D-3F)+2\sqrt{2} DS_\beta)/(2\sqrt{3} f_\pi)$\\
$\phi p \eta^\prime$& 0&$(-2\sqrt{2}D C_\beta +(D-3F)S_\beta)/(2\sqrt{3}f_\pi)$\\\\
\hline
$VBV^\prime$&$C_{V\to V^\prime P^\prime}$&$C_{B V^\prime\to B^\prime}$\\
&$\pi^0 p/\pi^0 N^{*+}$&$\pi^0 p$\\
\hline\\
$K^{*0}\Sigma^+ K^{*0}$&$-1/\sqrt{2}$&$-1$\\
$K^{*+}\Sigma^0 K^{*+}$&$1/\sqrt{2}$&$-1/\sqrt{2}$\\
$K^{*+}\Lambda K^{*+}$&$1/\sqrt{2}$&$-\sqrt{3/2}$\\
$\rho^0 p \rho^0$&$0$&$1/\sqrt{2}$\\
$\rho^0 p \omega$&$\sqrt{2}$&$3/\sqrt{2}$\\
$\rho^0 p \phi$&$0$&$0$\\
$\rho^+ n \rho^+$&$0$&$1$\\
$\omega p \rho^0$&$\sqrt{2}$&$1/\sqrt{2}$\\
$\omega p \omega$&$0$&$3/\sqrt{2}$\\
$\omega p \phi$&$0$&$0$\\
$\phi p \rho^0$& 0&$1/\sqrt{2}$\\
$\phi p \omega$& 0&$3/\sqrt{2}$\\
$\phi p \phi$& 0&$0$
\end{tabular}}
\end{table}

\begin{table}[h!]
\caption{Same as in Table~\ref{T1} but for the $\pi^+n$, and $\pi^+ N^{*0}$ final states. Here $N^{*0}$ represents either $N^{*0}(1535)$ or $N^{*0}(1650)$. In the case of $P_s\to \pi^+ N^{*0}$, $C_{BP\to B^\prime}=g_{N^{*0}\to PB}$, $C_{BV^\prime\to B^\prime}=g_{N^{*0}\to V^\prime B}$, i.e., the coupling constants of the resonance to the PB and VB channels.}
\resizebox{0.5\textwidth}{!}{
\begin{tabular}{ccc}
\hline\hline\\
$VBP$&$C_{V\to PP^\prime}$&$C_{B P\to B^\prime}$\\
&$\pi^+ n/\pi^+ N^{*0}$&$\pi^+ n$\\
\hline\\
$K^{*0}\Sigma^0 K^0$&$-1$&$(D-F)/(2f_\pi)$\\
$K^{*+}\Lambda K^0$&$-1$&$(D+3F)/(2\sqrt{3}f_\pi)$\\
$\rho^0 p \pi^-$&$-\sqrt{2}$&$-(D+F)/(\sqrt{2}f_\pi)$\\
$\rho^+ n \pi^0$&$\sqrt{2}$&$(D+F)/(2f_\pi)$\\
$\rho^+ n \eta$&$0$&$(C_\beta (D-3F)+2\sqrt{2} D S_\beta)/(2\sqrt{3}f_\pi)$\\
$\rho^+ n \eta^\prime$&$0$&$(-2\sqrt{2} D C_\beta +(D-3F) S_\beta)/(2\sqrt{3}f_\pi)$\\
$\omega p \pi^-$&$0$&$-(D+F)/(\sqrt{2}f_\pi)$\\
$\phi p \pi^-$& 0&$-(D+F)/(\sqrt{2}f_\pi)$\\
\hline
$VBV^\prime$&$C_{V\to V^\prime P^\prime}$&$C_{B V^\prime\to B^\prime}$\\
&$\pi^+ n/\pi^+ N^{*0}$&$\pi^+ n$\\
\hline\\
$K^{*0}\Sigma^0 K^{*0}$&$1$&$1/\sqrt{2}$\\
$K^{*+}\Lambda K^{*0}$&$1$&$-\sqrt{3/2}$\\
$\rho^0 p \rho^-$&$0$&$1$\\
$\rho^+ n \rho^0$&$0$&$-1/\sqrt{2}$\\
$\rho^+ n \omega$&$\sqrt{2}$&$3/\sqrt{2}$\\
$\rho^+ n \phi$&$0$&$0$\\
$\omega p \rho^-$&$\sqrt{2}$&$1$\\
$\phi p \rho^-$& 0&$1$\\
\end{tabular}}
\end{table}

\begin{table}[h!]
\caption{Same as in Table~\ref{T1} but for the $\eta p$, $\eta^\prime p$, and $\eta N^{*+}$ final states. In the case of $P_s\to \eta N^{*+}$, $C_{BP\to B^\prime}=g_{N^{*+}\to PB}$, $C_{BV^\prime\to B^\prime}=g_{N^{*+}\to V^\prime B}$, i.e., the coupling constants of the resonance to the PB and VB channels. Here $N^{*+}$ refers to either $N^{*+}(1535)$ or $N^{*+}(1650)$.}\label{Tf}
\resizebox{0.5\textwidth}{!}{
\begin{tabular}{cccc}
\hline\hline\\
$VBP$&\multicolumn{2}{c}{$C_{V\to PP^\prime}$}&$C_{B P\to B^\prime}$\\
&$\eta p/\eta N^{*+}$&$\eta^\prime p$&$\eta p/\eta^\prime p$\\
\hline\\
$K^{*0}\Sigma^+ K^0$&$-\sqrt{3/2}C_\beta$&$-\sqrt{3/2}S_\beta$&$(-D+F)/(\sqrt{2}f_\pi)$\\
$K^{*+}\Sigma^0 K^+$&$-\sqrt{3/2}C_\beta$&$-\sqrt{3/2}S_\beta$&$(-D+F)/(2f_\pi)$\\
$K^{*+}\Lambda K^+$&$-\sqrt{3/2}C_\beta$&$-\sqrt{3/2}S_\beta$&$(D+3F)/(2\sqrt{3}f_\pi)$\\
$\rho^0 p \pi^0$&$0$&$0$&$-(D+F)/(2f_\pi)$\\
$\rho^0 p \eta$&$0$&$0$&$(C_\beta (D-3F)+2\sqrt{2} D S_\beta)/(2\sqrt{3}f_\pi)$\\
$\rho^0 p \eta^\prime$&$0$&$0$&$(-2\sqrt{2}D C_\beta+(D-3F)S_\beta)/(2\sqrt{3}f_\pi)$\\
$\rho^+ n \pi^+$&$0$&$0$&$-(D+F)/(\sqrt{2}f_\pi)$\\
$\omega p \pi^0$&$0$&$0$&$-(D+F)/(2 f_\pi)$\\
$\omega p \eta$&$0$&$0$&$(C_\beta (D-3F)+2\sqrt{2} D S_\beta)/(2\sqrt{3}f_\pi)$\\
$\omega p \eta^\prime$&$0$&$0$&$(-2\sqrt{2}D C_\beta+(D-3F)S_\beta)/(2\sqrt{3}f_\pi)$\\
$\phi p \pi^0$& 0&$0$&$-(D+F)/(2 f_\pi)$\\
$\phi p \eta$& 0&$0$&$(C_\beta (D-3F)+2\sqrt{2} D S_\beta)/(2\sqrt{3}f_\pi)$\\
$\phi p \eta^\prime$& 0&$0$&$(-2\sqrt{2}D C_\beta+(D-3F)S_\beta)/(2\sqrt{3}f_\pi)$\\\\
\hline
$VBV^\prime$&\multicolumn{2}{c}{$C_{V\to V^\prime P^\prime}$}&$C_{B V^\prime\to B^\prime}$\\

&$\eta p/\eta N^{*+}$&$\eta^\prime p$&$\eta p/\eta^\prime p$\\
\hline\\
$K^{*0}\Sigma^+ K^{*0}$&$-(C_\beta+2\sqrt{2}S_\beta)/\sqrt{6}$&$(2\sqrt{2} C_\beta-S_\beta)/\sqrt{6}$&$-1$\\
$K^{*+}\Sigma^0 K^{*+}$&$-(C_\beta+2\sqrt{2}S_\beta)/\sqrt{6}$&$(2\sqrt{2} C_\beta-S_\beta)/\sqrt{6}$&$-1/\sqrt{2}$\\
$K^{*+}\Lambda K^{*+}$&$-(C_\beta+2\sqrt{2}S_\beta)/\sqrt{6}$&$(2\sqrt{2} C_\beta-S_\beta)/\sqrt{6}$&$-\sqrt{3/2}$\\
$\rho^0 p \rho^0$&$(\sqrt{2}C_\beta-2S_\beta)/\sqrt{3}$&$(2C_\beta+\sqrt{2}S_\beta)/\sqrt{3}$&$1/\sqrt{2}$\\
$\rho^0 p \omega$&$0$&$0$&$3/\sqrt{2}$\\
$\rho^0 p \phi$&$0$&$0$&$0$\\
$\rho^+ n \rho^+$&$(\sqrt{2}C_\beta-2S_\beta)/\sqrt{3}$&$(2C_\beta+\sqrt{2}S_\beta)/\sqrt{3}$&$1$\\
$\omega p \rho^0$&$0$&$0$&$1/\sqrt{2}$\\
$\omega p \omega$&$(\sqrt{2}C_\beta-2S_\beta)/\sqrt{3}$&$(2C_\beta+\sqrt{2}S_\beta)/\sqrt{3}$&$3/\sqrt{2}$\\
$\omega p \phi$&$0$&$0$&$0$\\
$\phi p \rho^0$& 0&$0$&$1/\sqrt{2}$\\
$\phi p \omega$& 0&$0$&$3/\sqrt{2}$\\
$\phi p \phi$& $-2(\sqrt{2}C_\beta+S_\beta)/\sqrt{3}$&$2(C_\beta-\sqrt{2}S_\beta)/\sqrt{3}$&$0$
\end{tabular}}
\end{table}
\clearpage

\section{Coefficients $F^{(l)}$, $H^{(l)}$, $J^{(l)}$, $\cdots$, $N^{(l)}$}\label{FHJ}
The coefficients $F^{(l)}$, $H^{(l)}$, $J^{(l)}$, $L^{(l)}$, $M^{(l)}$ and $N^{(l)}$ appearing in Eqs.~(\ref{Tr1}) and (\ref{Tr2}) can be written in terms of the coefficients $A_i$, $B_i$, $C_i$ and $D_i$ as follows:
\begin{align}
&F^{(l)}=\pmb{c}^{(l)}_F\cdot\pmb{v}_F;~H^{(l)}=\pmb{c}^{(l)}_H\cdot\pmb{v}_B;~J^{(l)}=\pmb{c}^{(l)}_J\cdot\pmb{v}_J;\nonumber\\
&L^{(l)}=\pmb{c}^{(l)}_L\cdot\pmb{v}_L;~M^{(l)}=\pmb{c}^{(l)}_M\cdot\pmb{v}_M;~N^{(l)}=\pmb{c}^{(l)}_N\cdot\pmb{v}_N,
\end{align}
where $\pmb{v}_F$, $\pmb{v}_B$, $\dots$, $\pmb{v}_N$ are vectors whose elements are:
\begin{align}
\pmb{v}_F&=\{|A_1|^2,2\text{Re}(A_1 A^*_3),2\text{Re}(A_1 A^*_4),2\text{Re}(A_1 A^*_7)\nonumber\\
&\quad\quad,2\text{Re}(A_1 \tilde{A}^*_3),2\text{Re}(A_1 \tilde{A}^*_4),|A_3|^2,2\text{Re}(A_3 A^*_4)\nonumber\\
&\quad\quad, 2\text{Re}(A_3 A^*_7),2\text{Re}(A_3 \tilde{A}^*_3),2\text{Re}(A_3 \tilde{A}^*_4),|A_4|^2\nonumber\\
&\quad\quad, 2\text{Re}(A_4 A^*_7),2\text{Re}(A_4 \tilde{A}^*_3),2\text{Re}(A_4 \tilde{A}^*_4),|A_7|^2\nonumber\\
&\quad\quad, 2\text{Re}(A_7 \tilde{A}^*_3),2\text{Re}(A_7 \tilde{A}^*_4),|\tilde{A}_3|^2,2\text{Re}(\tilde{A}_3 \tilde{A}^*_4)\nonumber\\
&\quad\quad, |\tilde{A}_4|^2\},\nonumber\\\nonumber\\
\pmb{v}_H&=\{|B_1|^2,2\text{Re}(B_1 B^*_2),2\text{Re}(B_1 B^*_3),2\text{Re}(B_1 B^*_4)\nonumber\\
&\quad\quad, 2\text{Re}(B_1 B^*_5), |B_2|^2,2\text{Re}(B_2 B^*_3),2\text{Re}(B_2 B^*_4)\nonumber\\
&\quad\quad, 2\text{Re}(B_2 B^*_5),|B_3|^2,2\text{Re}(B_3 B^*_4),2\text{Re}(B_3 B^*_5)\nonumber\\
&\quad\quad,|B_4|^2,2\text{Re}(B_4 B^*_5),|B_5|^2\},\nonumber\\\nonumber\\
\pmb{v}_J&=\{A_1 B^*_1,A_1 B^*_2,A_1 B^*_3,A_1 B^*_4,A_1 B^*_5,A_3 B^*_1\nonumber\\
&\quad\quad,A_3 B^*_2,A_3 B^*_3,A_3 B^*_4,A_3 B^*_5,A_4 B^*_1,A_4 B^*_2\nonumber\\
&\quad\quad,A_4 B^*_3,A_4 B^*_4,A_4 B^*_5,A_7 B^*_1,A_7 B^*_2,A_7 B^*_3\nonumber\\
&\quad\quad,A_7 B^*_4,A_7 B^*_5,\tilde{A}_3 B^*_1,\tilde{A}_3 B^*_2,\tilde{A}_3 B^*_3\nonumber\\
&\quad\quad,\tilde{A}_3 B^*_4,\tilde{A}_3 B^*_5,\tilde{A}_4 B^*_1,\tilde{A}_4 B^*_2,\tilde{A}_4 B^*_3,\tilde{A}_4 B^*_4\nonumber\\
&\quad\quad,\tilde{A}_4 B^*_5\}\nonumber\\\nonumber\\
\pmb{v}_L&=\{|C_3|^2,2\text{Re}(C_3 C^*_4),2\text{Re}(C_3 C^*_7),|C_4|^2\nonumber\\
&\quad\quad,2\text{Re}(C_4 C^*_7),|C_7|^2\},\nonumber\\\nonumber\\
\pmb{v}_M&=\{|D_1|^2,2\text{Re}(D_1 D^*_2),2\text{Re}(D_1 D^*_3),2\text{Re}(D_1 D^*_4)\nonumber\\
&\quad\quad,|D_2|^2,2\text{Re}(D_2 D^*_3),2\text{Re}(D_2 D^*_4),|D_3|^2\nonumber\\
&\quad\quad,2\text{Re}(D_3 D^*_4),|D_4|^2\},\nonumber\\\nonumber\\
\pmb{v}_N&=\{C_3 D^*_1,C_3 D^*_2,C_3 D^*_3,C_3 D^*_4,C_4 D^*_1,C_4 D^*_2\nonumber\\
&\quad\quad,C_4 D^*_3,C_4 D^*_4,C_7 D^*_1,C_7 D^*_2,C_7 D^*_3,C_7 D^*_4\}\nonumber.
\end{align}
In the following, we give some of the elements of the vectors $\pmb{c}^{(l)}_F$, $\pmb{c}^{(l)}_H$, $\dots$, $\pmb{c}^{(l)}_N$:
\begin{align}
\pmb{c}^{(0)}_F&=\Big\{-\frac{32}{3}m_{P_s}(m_{P_s}-m_{B^\prime})m^2_{P^\prime}, \frac{16}{3}(m_{B^\prime}-2m_{P_s})\nonumber\\
&\quad\quad\times m_{P_s}m^4_{P^\prime},-\frac{16}{3}m^2_{P_s}(m^2_{B^\prime}+m_{P_s}m_{B^\prime}-2m^2_{P_s})\nonumber\\
&\quad\quad\times m^2_{P^\prime},\dots\Big\},\nonumber\\
\pmb{c}^{(1)}_F&=\Big\{\frac{32}{3}m^2_{P^\prime},-\frac{16}{3}m^2_{P^\prime}[m^2_{B^\prime}+m_{P_s}m_{B^\prime}-2(m^2_{P_s}+m^2_{P^\prime})]\nonumber\\
&\quad\quad,\frac{16}{3}(m_{B^\prime}-4 m_{P_s})m_{P_s}m^2_{P^\prime},\dots\Big\},\nonumber\\
\pmb{c}^{(2)}_F&=\Big\{-\frac{32(m_{B^\prime}-m_{P_s})}{3m_{P_s}},-\frac{16m_{B^\prime}m^2_{P^\prime}}{3m_{P_s}}\nonumber\\
&\quad\quad,\frac{16}{3}(m^2_{B^\prime}+m_{P_s}m_{B^\prime}-2m^2_{P_s}+2m^2_{P^\prime}),\dots\Big\},\nonumber\\
\pmb{c}^{(3)}_F&=\Big\{-\frac{32}{3m^2_{P_s}},\frac{16[m^2_{B^\prime}+m_{P_s}m_{B^\prime}-2(m^2_{P_s}+m^2_{P^\prime})]}{3m^2_{P_s}}\nonumber\\
&\quad\quad,-\frac{16(m_{B^\prime}-4m_{P_s})}{3m_{P_s}},\dots\Big\},\nonumber\\
\pmb{c}^{(4)}_F&=\Big\{0,\frac{32}{3m^2_{P_s}},-\frac{32}{3m^2_{P_s}},\dots\Big\},\nonumber\\
\pmb{c}^{(5)}_F&=\Big\{0,0,0,0,0,0,-\frac{32}{3m^2_{P_s}},\frac{32}{3m^2_{P_s}},\dots\Big\},\nonumber\\
\pmb{c}^{(0)}_H&=\Big\{\frac{32}{3}m_{P_s}m^2_{P^\prime}(m_{B^\prime}-m_{P_s})\nonumber\\
&\quad\quad,\frac{16}{3}m^3_{P_s}m^2_{P^\prime}(m_{B^\prime}-m_{P_s}),\frac{16}{3}m^2_{P_s}m^4_{P^\prime},\dots\Big\},\nonumber\\
\pmb{c}^{(1)}_H&=\Big\{\frac{32 m^2_{P^\prime}}{3},\frac{16 m^2_{P_s}m^2_{P^\prime}}{3},\frac{16 m_{P_s}m^2_{P^\prime}(m_{B^\prime}-m_{P_s})}{3},\dots\Big\},\nonumber\\
\pmb{c}^{(2)}_H&=\Big\{-\frac{32(m_{B^\prime}-m_{P_s})}{3m_{P_s}},-\frac{16m_{P_s}(m_{B^\prime}-m_{P_s})}{3}\nonumber\\
&\quad\quad,-\frac{16m^2_{P^\prime}}{3},\dots\Big\},\nonumber\\
\pmb{c}^{(3)}_H&=\Big\{-\frac{32}{3m^2_{P_s}},-\frac{16}{3},-\frac{16(m_{B^\prime}-m_{P_s})}{3m_{P_s}},\dots\Big\},\nonumber\\
\pmb{c}^{(4)}_H&=\Big\{0,0,0,0,0,0,0,0,0,\frac{16}{3},\dots\Big\},\nonumber\\
\pmb{c}^{(0)}_J&=\Big\{-\frac{32m_{P_s}m^2_{P^\prime}(m_{P_s}-m_{B^\prime})}{3}\nonumber\\
&\quad\quad,\frac{16}{3}m^3_{P_s}m^2_{P^\prime}(m_{B^\prime}-m_{P_s}),\frac{16}{3}m^2_{P_s}m^4_{P^\prime},\dots\Big\},\nonumber\\
\pmb{c}^{(1)}_J&=\Big\{\frac{32m^2_{P^\prime}}{3},\frac{16 m^2_{P_s}m^2_{P^\prime}}{3}\nonumber\\
&\quad\quad,-\frac{16}{3}m_{P_s}m^2_{P^\prime}(m_{P_s}-m_{B^\prime}),\dots\Big\},\nonumber\\
\pmb{c}^{(2)}_J&=\Big\{-\frac{32(m_{B^\prime}-m_{P_s})}{3m_{P_s}},-\frac{16}{3}m_{P_s}(m_{B^\prime}-m_{P_s})\nonumber\\
&\quad\quad,-\frac{16}{3}m^2_{P^\prime},\dots\Big\},\nonumber\\
\pmb{c}^{(3)}_J&=\Big\{-\frac{32}{3m^2_{P_s}},-\frac{16}{3},-\frac{16(m_{B^\prime}-m_{P_s})}{3m_{P_s}},\dots\Big\},\nonumber\\
\pmb{c}^{(4)}_J&=\Big\{0,0,0,0,0,\frac{32}{3m^2_{P_s}},\frac{16}{3},\dots\Big\},\nonumber
\end{align}
\begin{align}
\pmb{c}^{(0)}_L&=\Big\{\frac{8m_{P_s}m^4_{P^\prime}(m_{P_s}-m_R)}{3},\frac{8m^2_{P_s}m^4_{P^\prime}}{3}\nonumber\\
&\quad\quad,\frac{8 m_{P_s}m^4_{P^\prime}}{3},\dots\Big\},\nonumber\\
\pmb{c}^{(1)}_L&=\Big\{\frac{8m^4_{P^\prime}}{3},-\frac{8m_{P_s}m^2_{P^\prime}(m_{P_s}+m_R)}{3}\nonumber\\
&\quad\quad,-\frac{8}{3}m^2_{P^\prime}(m_{P_s}+m_R),\dots\Big\},\nonumber\\
\pmb{c}^{(2)}_L&=\Big\{\frac{8}{3}\Big(\frac{m_R}{m_{P_s}}-3\Big)m^2_{P^\prime},-\frac{8m^2_{P^\prime}}{3},-\frac{8m^2_{P^\prime}}{3m_{P_s}},\dots\Big\},\nonumber\\
\pmb{c}^{(3)}_L&=\Big\{-\frac{8 m^2_{P^\prime}}{3m^2_{P_s}},\frac{8(m_{P_s}+m_R)}{3m_{P_s}},\frac{8(m_{P_s}+m_R)}{3m^2_{P_s}},\dots\Big\},\nonumber\\
\pmb{c}^{(4)}_L&=\Big\{\frac{16}{3m^2_{P_s}},0,0,\dots\Big\},\nonumber\\
\pmb{c}^{(0)}_M&=\Big\{-\frac{32m_{P_s}(m_{P_s}+m_R)m^2_{P^\prime}}{3}\nonumber\\
&\quad\quad,-\frac{16m^3_{P_s}m^2_{P^\prime}(m_{P_s}+m_R)}{3},\frac{16m^2_{P_s}m^4_{P^\prime}}{3},\dots\Big\}\nonumber\\
\pmb{c}^{(1)}_M&=\Big\{\frac{32 m^2_{P^\prime}}{3},\frac{16 m^2_{P_s}m^2_{P^\prime}}{3}\nonumber\\
&\quad\quad ,-\frac{16m_{P_s}m^2_{P^\prime}(m_{P_s}+m_R)}{3},\dots\Big\},\nonumber\\
\pmb{c}^{(2)}_M&=\Big\{\frac{32(m_{P_s}+m_R)}{3m_{P_s}},\frac{16m_{P_s}(m_{P_s}+m_R)}{3}\nonumber\\
&\quad\quad,-\frac{16 m^2_{P^\prime}}{3},\dots\Big\},\nonumber\\
\pmb{c}^{(3)}_M&=\Big\{-\frac{32}{3m^2_{P_s}},-\frac{16}{3},\frac{16(m_{P_s}+m_R)}{3m_{P_s}},\dots\Big\},\nonumber\\
\pmb{c}^{(4)}_M&=\Big\{0,0,0,0,0,0,0,\frac{16}{3},\dots,\Big\},\nonumber
\end{align}

\begin{align}
\pmb{c}^{(0)}_N&=\Big\{\frac{16m^4_{P^\prime}m_{P_s}}{3},\frac{8m^4_{P^\prime}m^3_{P_s}}{3}\nonumber\\
&\quad\quad,\frac{8 m^4_{P^\prime}m^2_{P_s}(m_{P_s}-m_R)}{3},\dots\Big\},\nonumber\\
\pmb{c}^{(1)}_N&=\Big\{-\frac{16 m^2_{P^\prime}(m_{P_s}+m_R)}{3}\nonumber\\
&\quad\quad,-\frac{8m^2_{P^\prime}m^2_{P_s}(m_{P_s}+m_R)}{3},\frac{8m^4_{P^\prime}m_{P_s}}{3},\dots\Big\},\nonumber\\
\pmb{c}^{(2)}_N&=\Big\{-\frac{16 m^2_{P^\prime}}{3 m_{P_s}},-\frac{8m^2_{P^\prime}m_{P_s}}{3}\nonumber\\
&\quad\quad,-\frac{8m^2_{P^\prime}(3m_{P_s}-m_R)}{3},\dots\Big\},\nonumber\\
\pmb{c}^{(3)}_N&=\Big\{\frac{16(m_{P_s}+m_R)}{3m^2_{P_s}},\frac{8(m_{P_s}+m_R)}{3}\nonumber\\
&\quad\quad,-\frac{8m^2_{P^\prime}}{3m_{P_s}},\dots\Big\},\nonumber\\
\pmb{c}^{(4)}_N&=\Big\{0,0,\frac{16}{3m_{P_s}},0,\dots\Big\}.
\end{align}

\clearpage
\bibliographystyle{unsrt}
\bibliography{refs}

\end{document}